\documentclass[onecolumn]{elsart3p}

\usepackage{graphicx}

\usepackage{amssymb}
\usepackage{amsmath}
\usepackage{color}
\usepackage[english,francais]{babel}
\usepackage{enumerate}

\newtheorem{e-proposition}[theorem]{Proposition}

\newtheorem{e-definition}[theorem]{Definition\rm}

\renewcommand{\theequation}{\arabic{equation}}
\def\og{\leavevmode\raise.3ex\hbox{$\scriptscriptstyle\langle\!\langle$~}}
\def\fg{\leavevmode\raise.3ex\hbox{~$\!\scriptscriptstyle\,\rangle\!\rangle$}}

\journal{CR Physique}

\def\xhat{\hat{\mathbf{x}}}
\def\yhat{\hat{\mathbf{y}}}

\def\ket#1{{\left|#1\right\rangle}}
\def\bra#1{{\left\langle #1 \right|}}

\def\br{\mathbf{r}}
\def\bk{\mathbf{k}}
\def\bq{\mathbf{q}}

\def\bR{\mathbf{R}}

\def\Winf{W_\infty}

\def\be{\begin{eqnarray}}
\def\ee{\end{eqnarray}}

\begin{document}
\centerline{Title of the dossier/Titre du dossier}
\begin{frontmatter}


\selectlanguage{english}
\title{
Fractional Quantum Hall Physics in Topological Flat Bands
}


\selectlanguage{english}
\author{S. A. Parameswaran}
\address{Department of Physics, University of California, Berkeley, California 94720}
\ead{sidp@berkeley.edu, {\rm Corresponding Author}}
\author{R. Roy}
\address{Department of Physics and Astronomy, University
of California, Los Angeles, California 90095-1547}            
\ead{rroy@physics.ucla.edu}
\author{S. L. Sondhi}
\address{Department of Physics, Princeton
University, Princeton, NJ 08544}     
\ead{sondhi@princeton.edu}


\begin{abstract}
We present a pedagogical review of the physics of fractional Chern insulators with a particular focus on the connection to the fractional quantum Hall effect. While the latter conventionally arises in semiconductor heterostructures at low temperatures and in high magnetic fields, interacting Chern insulators at fractional band filling may host phases with the same topological properties, but stabilized at the lattice scale, potentially leading to high-temperature topological order. We discuss the construction of topological flat band models, provide a survey of numerical results, and establish the connection between the Chern band and the continuum Landau problem. We then briefly summarize various aspects of Chern band physics that have no natural continuum analogs, before turning to a discussion of possible experimental realizations. We close with a survey of future directions and open problems, as well as a discussion of extensions of these ideas to higher dimensions and to other topological phases.
\vskip 0.5\baselineskip

\selectlanguage{francais}
\noindent{\bf R\'esum\'e}
\vskip 0.5\baselineskip
\noindent
{\bf Here is the French title. }
Your r\'esum\'e in French here.

 \vskip 0.5\baselineskip
\noindent{\small{\it Mots-cl\'es~:} Mot-cl\'e1~; Mot-cl\'e2~;
Mot-cl\'e3}}
\end{abstract}
\end{frontmatter}

\selectlanguage{english}

\noindent

\section{Introduction}
The application of a transverse magnetic field to a two-dimensional electron gas breaks time-reversal symmetry and gives rise to a highly degenerate single-particle spectrum in the absence of electron-electron interections: the celebrated Landau levels.  When the electronic density is commensurate with the macroscopic degeneracy of a Landau level so that an integer number of such levels are occupied, the resulting insulating phase exhibits the integer quantum Hall effect (IQHE), with spectacular consequences for charge transport: the Hall resistance is quantized to be an integer multiple of $h/e^2$, while the longitudinal resistance vanishes \cite{vonKlitzing:1980p1}. When the density is a rational fraction $p/q<1$ of the Landau level degeneracy, there is no appropriate single-particle description and it lies to interactions to select a suitable correlated ground state from the degenerate manifold. The resulting phases exhibit the fractional quantum Hall effect (FQHE) \cite{Tsui:1982p1,laughlin1983anomalous}. The FQH phases are prototypical example of topologically ordered \cite{Wen:1990p1} phases of matter: they are described at low energies by a topological quantum field theory written in terms of an emergent gauge field, exhibit fractionalization of quantum numbers in the bulk, and have gapless protected chiral edge states. 

The simplicity of the setting in which the QHE was first discovered was extremely helpful in unraveling its explanation \cite{Laughlin:1983p1,laughlin1983anomalous}. For the two-dimensional electron gases (2DEGs) in semiconductor heterostructures that are the venue for all QH experiments thus far, we may ignore most of the typical complications associated with the surrounding solid and treat the electrons as having a parabolic dispersion. Lattice effects at best lead to a change in the (effective) mass of electrons away from its value in free space. Hand-in-hand with this description in terms of nearly free electrons is the fact that the structure of single particle states in the lowest Landau level (henceforth abbreviated as LLL) strongly constrains the wave functions of the many-body ground states that emerge as a consequence of interactions: they are forced to be {\it analytic} functions of $z=x+iy$. The built-in analyticity of the LLL wave functions and the structure of the topological quantum field theory that describes their topological order has allowed the full panoply of techniques of conformal field theory to be brought to bear in constructing variational wave functions for various QHE `filling fractions', $\nu =p/q$ \cite{laughlin1983anomalous,Haldane:1983p1,Halperin:1984p1,MOORE:1991p58}. Various effective field theory techniques, such as statistical transmutation by flux attachment, \cite{ZHANG:1989p1367,zhang1992chern} also rely on the simple parabolic dispersion assumed for the electrons.

Can the QHE be detached from this idealized limit? There are two independent idealizations that must be addressed. First,  the solid could affect electronic motion more seriously so that the effective mass approximation gives way to the formation of energy bands; does the QHE survive in this case? Second, a uniform magnetic field both breaks time reversal symmetry but also affects electron dynamics at long wavelengths in a decidedly unusual fashion: the formation of Landau levels. Are both essential for the QHE?

The answer to both questions is known for the I(nteger)QHE, which is mostly a single-particle phenomenon. In a landmark paper in 1982, Thouless, Kohmoto, Nightingale, and den Nijs \cite{TKNN} analyzed the uniform-field Hall effect in a strong periodic potential that was known to lead to an intricate spectrum, the so-called Hofstadter butterfly; they showed that it gives rise to an integer QHE under certain conditions, i.e., whenever the chemical potential lies in a gap. Indeed, the Hall conductance was shown to map to a topological invariant associated with filled bands --- the (first) Chern number. Six years later, in another striking development, Haldane \cite{Haldane:1988p1} answered the second question, showing by an explicit construction of a tight-binding model on a honeycomb lattice that a quantized Hall conductance can arise from a fully filled band even in the absence of a net magnetic field. In his model, time-reversal symmetry is broken by a spatially inhomogeneous magnetic field with 
zero average, and the Hall conductance again equals the Chern number of the band.

Of course, the next logical question is: can the FQHE, canonically a property of interacting electrons in a fractionally filled Landau level, also be separated from the weak lattice and uniform magnetic field limit? The topological flat band models attempt to ask this question in a specific fashion: if it is true for independent electrons that a filled Chern band is equivalent to a filled Landau level, then is it also true for interacting electrons that a fractionally filled Chern band is equivalent to a fractionally filled Landau level? Microscopically, these models proceed to construct a (nearly) degenerate low-energy  subspace -- the `flat band'. While an exact degeneracy requires electron hopping over arbitrarly large distances, the hopping amplitudes decay exponentially and thus a relatively flat band can be produced by keeping a small set of hopping amplitudes. The suitability of a band for realizing correlated states is na\"ively quantified by its flatness parameter: the ratio of the band gap (which sets a bound on the strength of the interactions one can safely include) to the bandwidth -- although, as we will see, this is only one criterion for a `good' Chern band.

To understand the significance of a lattice realization of the FQHE it is useful to recall that while in theoretical terms the setting in which the FQHE was discovered is quite simple, experiments must go to great lengths to achieve a limit where this simplicity emerges. First, a variety of growth techniques must be employed to confine electrons in a two-dimensional plane,while simultaneously maintaining a high electron mobility -- the latter restriction because too much disorder destroys the FQHE. Second, an extremely strong magnetic field -- of the order of tens of tesla -- must be applied for Landau quantization to be appreciable. Finally, the systems must be cooled to below at least $10$ K to reach the energy scales at which FQH physics is manifest. All in all, this involves a {\it tour de force} in experimental technique, and reliably reproducing the full sequence of FQH states remains a significant challenge even three decades after the original discovery. In contrast, in a lattice model, the characteristic energy scales can be in the tens or even hundreds of kelvin (assuming a lattice energy of a few meV) and are naturally in the limit where each unit cell sees a large effective magnetic field.  Therefore such systems may realize robust phases that are concomitantly insensitive to disorder and thermal fluctuations. The possibility of stabilizing exotic topological phases outside of a dilution fridge and without a superconducting magnet, and the avenues it opens for further experimental probes of these phases, is among the primary motivations of the activity in this field. On a more theoretical front, the interplay of lattice symmetries with topological order leads to new physics, unique to the lattice realizations of the FQHE.

The recent flurry of interest in the FQHE in topological flat bands began with the work of three groups \cite{Tang:2011p1,Sun:2011p1,Neupert:2011p1}; several authors have since constructed models with nearly flat (non-dispersing) Chern bands which allow interactions to dominate at partial fillings while
leaving the gap to neighboring bands open. Initial papers \cite{Neupert:2011p1,sheng2011fractional,PhysRevLett.108.126805}  reported evidence for FQH states at $\nu=1/3$, $1/5$ and (for bosons) $1/2$,  in finite-size studies of short-ranged interactions projected to these bands. A detailed study of properties of ground state entanglement \cite{regnault2011fractional} allows a rationalization of some aspects of these results in terms of a generalized Pauli principle familiar from the LLL problem. The connection to the FQHE  was made more explicit by Qi \cite{Qi:2011p1} who gave
a fairly general recipe for translating familiar model wave functions and Hamiltonians
from the LLL to Chern bands on cylinders by elegantly mapping Landau gauge eigenfunctions to particular
Wannier functions. Finally, work by the present authors \cite{parameswaran2012fractional,Roy2012Geometry} gave a mapping between Hamiltonians in the LLL and those in the Chern band by studying the correspondence of the operator algebras that arise in the two cases. The outcome of these three  early approaches is to identify three desiderata for a Chern band: (i) that it be { energetically flat} with a large gap and thus a large flatness parameter, so that the low-energy description may be approximated by projecting into the band (ii) that it be {\it Berry flat}, i.e. have near-uniform Berry curvature, so that the algebra of densities projected to the Chern band resembles that of similar operators in the LLL and (iii) that it satisfy a set of conditions on the Fubini-Study metric of band eigenstates that quantify the connection to the LLL. Building on this, a reformulation of the Hamiltonian theory of the FQHE has been applied to study Chern bands~\cite{murthy2011composite,2012arXiv1207.2133M}.

This is as good a place as any to note that on a lattice the distinction between having a net magnetic field and not having it at all is not as sharp as it may seem. Essentially, it is always possible to stick a full flux quantum through some subset of loops on the lattice to shift the average magnetic field without affecting the actual physics. From this perspective, the physics in these flat-band models has a family resemblance to earlier studies of lattice versions of the FQHE \cite{Sorenson:2005p1,Moller:2009p2}  with uniform magnetic fields. In this earlier work, the authors studied a fixed filling factor while varying the flux per plaquette from small values and large unit cells, where the standard Landau level description holds, to somewhat larger flux values and smaller unit cells, where that description broke down. As they were able to change this parameter without any evidence of encountering a phase transition, the latter limit constituted an observation of the FQHE in the presence of strong lattice effects. As we demonstrate below, a more analytic approach clarifies the equivalence between this earlier ÒHofstadterÓ and the current ÒHaldaneÓ versions of the FQHE. Furthermore, this connection permits the construction of closed algebra of one-body operators \cite{parameswaran2012fractional,Roy2012Geometry,murthy2011composite,2012arXiv1207.2133M} that gives a Chern band realization of the Girvin-Macdonald-Platzmann algebra \cite{GMP:1986p1} of the LLL.

In addition, several authors have examined aspects of Chern band physics where the presence of the lattice introduces an additional layer of richness and complexity.  We discuss two examples in this review. The first of these is the intriguing question of the transition from a kinetic-energy dominated regime where the inter-particle interactions are weak and the system is in a gapless metallic phase (superfluid, if bosonic) to the strong-coupling regime where the flat band FQHE sets in. {\it Continuous} bandwidth-tuned transitions between these two limits provide instances of topological phase transitions \cite{PhysRevB.86.075136,Barkeshli12014393}, which we discuss briefly below.
A second example of the reappearance of lattice physics is the case of bands with higher Chern number. Closely related to `bilayer' quantum Hall states in a Landau level, higher Chern bands both provide a simple route to non-abelian FQH states \cite{2012arXiv1209.2424Z}. In addition, the lattice physics reappears via a nontrivial interplay of crystalline point defects with the topological order, endowing them  with a  nontrivial `quantum dimension.' 

At the time of writing, FQH physics in Chern bands is numerically well-tested, with the proposed candidates having passed a host of detailed checks by various groups. However, an experimental realization of a fractional Chern band remains elusive, although several candidates have been 
suggested. We will discuss a few promising ones below: these include solid-state realizations in oxide interfaces \cite{Xiao:2011fk}, as well as optical lattice examples both with short-range \cite{PhysRevLett.106.175301} as well as dipolar interactions \cite{2012arXiv1212.4839Y,PhysRevLett.109.266804}.

\subsection{This Review}
Before proceeding, we list a few disclaimers as to the scope of this review. As with any rapidly progressing field, there is a risk that some of what we discuss will be out of date by the time of publication, or that currently active directions fail to realize their early promise. To avoid this problem of obsolescence and in the interests of pedagogy our primary focus is on the case where the characterization at the present time is most complete:  Chern bands in two dimensions. We will eschew a discussion of other topological bands --- such as those relevant to fractionalized analogs of time reversal invariant $\mathbb{Z}_2$ topological insulators --- and higher dimensions. The bulk of the review will focus on bands with Chern number $C=1$, although we will briefly discuss bands with $C=2$. We will also restrict our detailed discussion to the early work on the subject, and devote a significant fraction of this review to elucidating the relationship between FQHE states realized in the continuum, and their putative analogs in Chern bands. We will attempt to present a broad summary of the subsequent direction of the field, with pointers to the literature for the interested reader.  We will also briefly discuss new aspects of FQHE physics unique to Chern bands, as well as potential experimental realizations. Also, while we describe the important numerical results, we will not discuss the specifics of the numerics in this review, referring the reader to the original literature for further details.  Our intention is that by the end of this article, a novice to the subject will emerge familiar with the broader aspects of fractionally filled topological bands, and equipped to tackle the (rapidly growing) literature; we trust that the reader will not find our approach too idiosyncratic.

\section{Survey of Models}
We begin with a brief introduction to Chern insulators in general and flat band models in particular, that will serve both to fix notation and provide background for the remainder of this review. Following a general discussion of tight-binding models without interactions, we proceed to the question of interacting flat band models where we review the numerical results that initiated the systematic study of fractional Chern insulators. We also briefly discuss measures such as ground state entanglement, that can be unambiguously used to identify the topologically ordered phases in these models. 

\subsection{Lattice Models of Chern Insulators}
Consider a tight-binding lattice model of the form 
\be
H_0 = \sum_{i,j, a,b} t^{ab}_{ij}c^\dagger_{i,a} c_{j,b},\ee
 where the $\{t_{ij}^{ab}\}$ are the (in general complex) hopping matrix elements, the $i,j$ label Bravais lattice sites,  and  $a,b =1,2,\ldots,\mathcal{N}$ are internal indices that label different orbitals or sites  within a unit cell. Invoking Bloch's theorem and working in momentum space, we may write the Hamiltonian as
\be
H_0 = \sum_{\bk,a,b} c^\dagger_{\bk,a} h_{ab}(\bk) c_{\bk,b}
\ee
where $a,b$ are 
indices that label sites within a unit cell, and 
 $\bk$ is the crystal momentum restricted to the first Brillouin zone (BZ). (Here and below, we will explicitly indicate  when repeated indices are summed over, and unless otherwise specified will work in two dimensions.)
The solution of the $\mathcal{N}\times\mathcal{N}$ eigenvalue problem
$\sum_{b} h_{ab}(\bk) u^{\alpha}_b(\bk) = \epsilon_{\alpha}(\bk) u^{\alpha}_a(\bk)$
defines the Bloch bands $\epsilon_\alpha(\bk)$ and Bloch states $u^\alpha_a(\bk)$. We will take the corresponding eigenvectors to be normalized, $\sum_{a} |u^{\alpha}_{a}(\bk)|^2 =1$.
The corresponding eigenstates are given by
\be\label{eq:eigstate}
\ket{\bk,\alpha} = \gamma^\dagger_{\bk,\alpha}\ket{0} &\equiv& \sum_a u^{\alpha}_{a}(\bk)c^\dagger_{\bk,a} \ket{0}
\ee
and in terms of the operators $\gamma^\dagger_{\bk,\alpha}$ we have $H_0 = \sum_{\alpha=1}^{\mathcal{N}} \epsilon_\alpha(\bk) \gamma^\dagger_{\bk,\alpha}\gamma_{\bk,\alpha}$. 

 The Chern number of a given band $\alpha$ is a topological invariant which can be defined only if the band is isolated from all other bands, and is computed as
\be\label{eq:Chernnumber}
C_\alpha = \frac{1}{2\pi} \int _{\text{BZ}}{d^2 k}\, \mathcal{B}_\alpha(\bk).\ee
Here, $\mathcal{B}_\alpha(\bk)$ is the  Chern flux density (Berry curvature), defined as the curl of the Berry connection (Berry gauge potential), $\mathcal{B}_\alpha(\bk) = \vec{\nabla}_\bk \times \vec{\mathcal{A}}_\alpha(\bk)$. In terms of the Bloch states, we have 
\be\label{eq:Chernconnection}
 \vec{\mathcal{A}}_\alpha(\bk) =  i \sum_{b=1}^\mathcal{N} u^{\alpha*}_b(\bk) \vec{\nabla}_\bk u^\alpha_b(\bk) .
\ee
A filled band with Chern number $C_\alpha$ yields a Hall conductance $\sigma_H = C_\alpha e^2/h$ regardless of whether it arises in a system with a net magnetic field \cite{TKNN}
(``Hofstadter band'') or zero net magnetic field \cite{Haldane:1988p1} (``Haldane band''). We shall 
refer to both as Chern bands.
In addition, we will assume that we are considering bands with $C_\alpha=1$ unless otherwise specified.

\subsection{Engineering Flat Bands}
A Chern band shares one important feature with a Landau level: a nonzero Chern number. A Landau level has the additional feature that it is an exactly degenerate manifold, so that the only energy scale within a fractionally filled Landau level is provided by the interparticle interactions. In a typical Chern band, however, the kinetic energy from the band dispersion is a competing energy scale. In order to increase the efficacy of interactions the kinetic energy can be `quenched' by flattening the dispersion, as was noted by several authors; it is important that the single-particle Hamiltonian remain local even after the band-flattening procedure.  Consider a lattice model of the form above, and let us assume the lowest band\footnote{The assumption of the lowest band is for pedagogical reasons; the generalization to the case when there are several filled bands below the one of interest is trivial and left as an exercise to the reader.} has Chern number $1$ and gap 
$\Delta =  \min(\epsilon_{2}(\bk)) - \max (\epsilon_{1}(\bk))$. 
Simply performing the transformation 
\be h_{ab}(\bk)\rightarrow \tilde{h}_{ab}(\bk) = h_{ab}(\bk)/\epsilon_1(\bk)
\ee 
and Fourier transforming to return to real space defines a tight-binding model in which the hopping matrix elements decay asymptotically as $\tilde{t}_{ij}^{ab} \propto e^{ -|i-j|/\Delta}$ and the lowest band is flat. Thus, as a point of principle, {\it any} energy band that is gapped away from all other bands can be made flat, at the cost of adding exponentially decaying longer-range hoppings; while the resulting tight-binding model is not strictly local (in the sense of having matrix elements with compact support), most arguments based on topological order remain valid for the exponentially local Hamiltonian that results in this fashion. Of course, truncating the hopping elements for sufficiently large $|i-j|$ renders the Hamiltonian strictly local at the cost of introducing some dispersion, characterized by a bandwidth $t$. Correlations then play the dominant role only if the interaction scale is much larger than the bandwidth, $U\gg t$. The maximum interaction scale is itself limited by the band gap, since the approximation of restricting attention to the lowest band breaks down unless $U\ll\Delta$. Therefore an appropriate quantitative measure of the suitability of a Chern band for realizing correlated phases isthe flatness parameter $f = \Delta/t$. Particles in bands with large $f$ have correlation energies much higher than their kinetic energies, for interacting Hamiltonians that may safely be approximated by their projection into the band. 
Several authors~\cite{PhysRevB.84.155116,Yang:PhysRevB86:2012,PhysRevB.84.241103,PhysRevB.86.241111} all constructed flat bands with fairly large flatness parameters, even with hopping restricted to a few near neighbors. This solves the first problem we noted earlier: we have successfully produced examples in which it is reasonable to approximate the physics by projecting to a single band with non-trivial topology.

\subsection{Interacting Flat Band Models}
Once a noninteracting Chern insulator hamiltonian $H_0$ has been obtained that hosts a flat band $\alpha$ with large flatness parameter, the next step is to add inter-particle interactions. The full interacting hamiltonian takes the form $
H = H_0 +V$ where $V$ is an interaction term, which typically (but not always) takes a generalized Hubbard form:
\be
V = \frac{1}{2}\sum_{i,j} U_{ij} \hat{n}_i \hat{n}_j
\ee
where $\hat{n}_i = c^\dagger_i c_i$ is the number operator for electrons on site $i$. For such a density-density interaction, assuming a translationally invariant Hamiltonian, we have $U_{ij}  = U(\br_i -\br_j) = \int \frac{d^2q}{(2\pi)^2} U(\bq) e^{i\bq\cdot(\br_i-\br_j)}$.  The interaction term can then be written $V = \sum_{\bq} U(q) \hat{n}_{\bq}\hat{n}_{-\bq}$ where $\hat{n}_\bq$ is the Fourier transform of $\hat{n}_i$. As a measure of the interaction scale $U$ we can take (for instance) the maximum value of $U(\br_i-\br_j)$. Note that the full space of interacting Hamiltonians is somewhat more general than the density-density interaction assumed above -- for instance it could include ring exchange, pair hopping and other non-density contributions. Nevertheless we will mostly restrict ourselves to such Hubbard-type models as they are a reasonable starting point and have a well-founded physical origin in an onsite Coulomb repulsion.

\subsection{Projecting to the Flat Band}
Since we assume $U\ll \Delta$ we may safely neglect the mixing between $\alpha$ and the remaining bands. This allows us to project to the partially filled band; such a projection is implemented by the operator $\mathcal{P}_\alpha = \sum_{\bk} \ket{\bk,\alpha}\bra{\bk,\alpha}$. In second-quantized form, the resulting Hamiltonian takes the form
\be
\overline{H} &=& \mathcal{P}_\alpha H \mathcal{P_\alpha} 
= \sum_{\bq} \epsilon_\alpha(\bq) \gamma^\dagger_{\bq,\alpha}\gamma_{\bq, \alpha}  +\frac{1}{2}\sum_{\bq} U(\bq) \bar{\rho}_{\bq;\alpha} \bar{\rho}_{-\bq;\alpha}
\ee
where we have defined the projected density operator in the Chern band,
\be\label{eq:CBprojdens}
\bar{\rho}_{\bq;\alpha} &=& \mathcal{P}_\alpha \hat{n}_\bq \mathcal{P_\alpha}  
=\sum_{\bk,b}
u^{\alpha *}_b\left(\bk+\frac{\bq}{2}\right) u^\alpha_b \left(\bk-\frac{\bq}{2}\right) 
\times \gamma^\dagger_{\bk+\frac{\bq}{2},\alpha}\gamma_{ \bk-\frac{\bq}{2},\alpha}
\ee
If the bandwidth
is small compared to the scale of the interactions, $t\ll U$, 
$\epsilon_{\alpha}(\bk)$ may be treated as constant and thus ignored. 
With this approximation, we finally arrive at the
low energy effective flat band Hamiltonian, which takes the form  
\be
H_{\text{CB},\alpha}^{\text{eff}}=  \frac{1}{2}\sum_{\bq} U(\bq) \bar{\rho}_{\bq;\alpha} \bar{\rho}_{-\bq;\alpha}
\ee

One encounters a similar Hamiltonian in the treatment of
interactions in the LLL in a large magnetic
field. In that case, the effective Hamiltonian of a clean
system obtained by projecting density-density interactions to the LLL has the form
\begin{align}
H^{\text{eff}}_{\text{LLL}} = \frac{1}{2} \sum_{\bq} V(\bq)e^{-q^2\ell_B^2/4}
\overline{\rho}_\bq \overline{\rho}_{-\bq}  \label{LLL-proj-int}
\end{align}
where in this case
$\overline{\rho}_{\bq}$ differs from the projected density, 
$\mathcal{P} \rho_\bq \mathcal{P}$ by a $q$-dependent
constant, $ \mathcal{P} \rho_\bq \mathcal{P} =
e^{-q^2\ell_B^2/4} e^{i\bq\cdot\bR} \equiv e^{-q^2\ell_B^2/2}
\overline{\rho}_{\bq}$ where  $\mathcal{P}$ is the operator that projects to the LLL, $\bR$ is the `guiding center' coordinate and $\ell_B = (\hbar c/ eB)$ is the magnetic length. $V(\bq)$ is once again the Fourier transform of a two-body density-density interaction.

The similarity between Hamiltonians projected to a Landau level and to a Chern band, coupled with the fact that both the LL and the Chern band have a nontrivial Chern number suggests that similar fractionalized phases may arise in the Chern band as in the FQHE.  The missing piece in making this analogy precise is to match the commutation relations of the projected densities, which we discuss at length below. First, we give a brief overview of the numerical studies that have driven much of the early progress in the field.

\section{Numerical Results}
Once a flat band Hamiltonian has been constructed, the next logical step is to ascertain the properties of its ground state at a given filling. Analogously to the study of the FQHE in the LLL, the approach of choice is to exactly diagonalize the flat band Hamiltonian for small numbers of electrons, and try and extract from this the ground state properties in the thermodynamic limit. Since such analysis is built from the existing intuition gleaned from LLL numerics, it is useful to first review those original aspects. We should preface this discussion by noting that while much of the numerical intuition for the LLL is obtained in the spherical geometry where even topologically ordered ground states are non-degenerate, here we discuss continuum results that study FQH states on a torus as they can be directly compared to lattice Chern bands.

On a torus, FQH states are expected to exhibit a ground state degeneracy in the thermodynamic limit: they should have a low-lying multiplet of topologically degenerate states, e.g. there are $m$ such levels for the $1/m$ Laughlin state.\footnote{However, it is possible that other states, with broken symmetries -- for instance those with charge density wave order--  also exhibit a similar degeneracy on a torus as first noted by Haldane \cite{PhysRevLett.55.2095} for the case of continuum LL problem, and so some care must be taken in using the ground state degeneracy alone to judge if the ground state is in a FQH phase.} The gap between these and the remaining states in the spectrum should scale to a constant in the thermodynamic limit. Furthermore, upon `flux insertion' through a handle of the torus -- implemented operationally by diagonalizing the problem with twisted boundary conditions -- the states should exhibit `spectral flow' within the low-lying multiplet: as the flux is increased, the levels should evolve into each other, with a periodicity in $m$ flux quanta for the $1/m$ Laughlin states. For each degenerate ground state, a many-body Hall conductance can also be computed by suitably twisting boundary conditions, and is expected to match that of the FQH phase. An additional test is to study the uniformity of the particle densities, as the FQH state is expected to be a uniform incompressible liquid. The overlap of a numerical ground state with trial wave functions --- which can  be analytically demonstrated to be in a particular FQH phase --  provides further characterization of its properties. Similarly, several of the FQH wave functions are gapped exact ground states of local model `pseudopotential Hamiltonians'; adiabatic continuity between these and the actual Hamiltonian can be used to establish the ground state topological order. The fractional statistics of the quasiparticle excitations of the FQH also follow a `generalized Pauli principle' that governs the structure of low-lying many-body eigenstates with a fixed number of excitations in a precise manner that can be tested numerically.
Finally, a more sophisticated tool  is to study ground state {\it entanglement}, whose properties can be rationalized in terms of topological field theoretic considerations and the bulk-edge correspondence of the FQHE.  Each of these approaches, to a greater or lesser degree, has been transcribed to the problem of fractionally filled Chern bands.

\subsection{First Results: Overlaps, Hall Conductances and Spectral Flow}
\begin{figure}\center
\includegraphics[width=0.4\columnwidth]{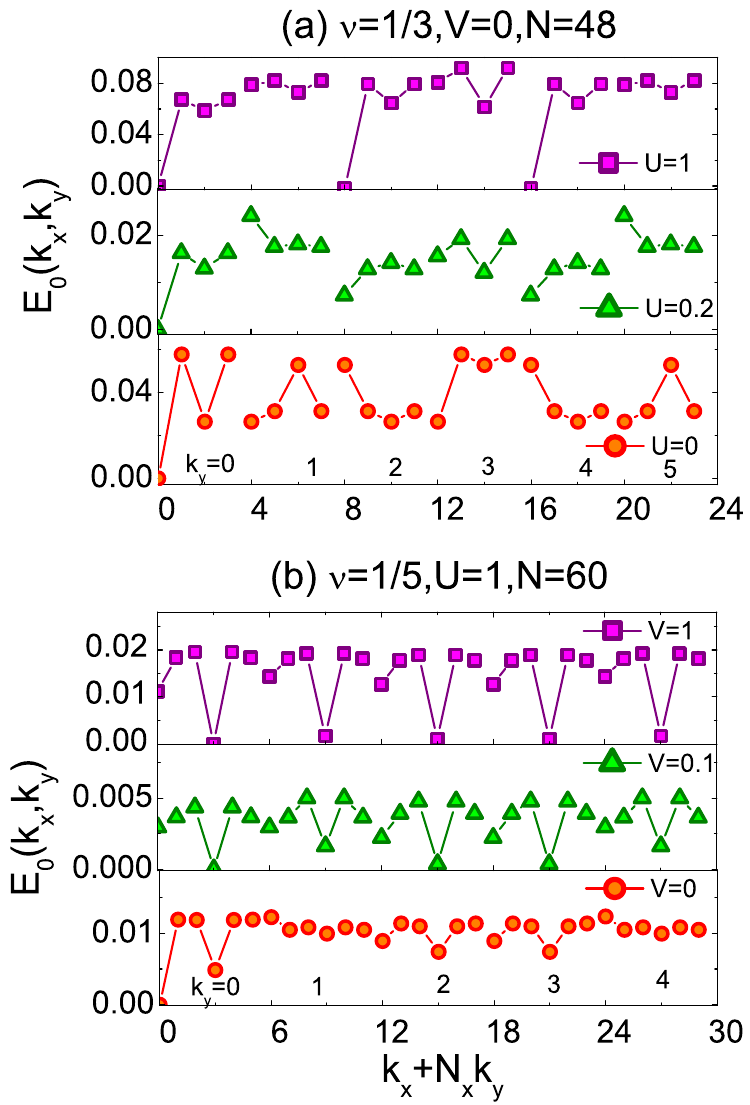}
\caption{{\bf Low energy spectrum of fractional Chern insulators and other phases in fractional Chern bands from Ref.~\cite{sheng2011fractional}}. The parameters U and V control the scale of the nearest neighbor and next-nearest neighbor repulsion terms in the Hamiltonian. The top panels in (a) and (b) clearly illustrate the difference in scale between the energy spread of the ground state multiplet and the energy gap. 
\label{Sheng figure}}
\end{figure}

The earliest evidence  for FQH physics took the form of studying ground state degeneracies as well as spectral flow, as well as a computation of ground state Hall conductance \footnote{We stress that this was at that point state-of-the-art, since the analytic understanding of the FQH in Chern bands was then insufficient to implement various other checks.} For instance, Neupert {\it et.~al.}~\cite{Neupert:2011p1}  studied systems with $6$-electrons at $1/3$ filling, and demonstrated spectral flow between the three lowest eigenstates and an absence of flow among higher levels. Although the gap between some of the putative ground states and the remainder of the spectrum was not particularly large and comparable to the splitting between the three lowest states, the spectral flow coupled with numerical estimates of the Hall conductance provided early evidence for topological order. Subsequent numerics by Sheng {\it et.~al.}~\cite{sheng2011fractional} --  on a slightly different 
 flat band model, slightly larger system sizes, and appropriately tuned interactions -- found gaps at fillings $1/3$ and $1/5$  which were much larger than the spread of the lowest lying multiplet~\ref{Sheng figure}. Furthermore, while exact parent Hamiltonians in the Chern band were still unknown, the Hamiltonians realizing these FQH phases were of the same general form (in terms of range of repulsion, etc.) as the Haldane pseudopotentials in the LLL -- for instance, to stabilize the $1/5$ state, next nearest neighbor interactions needed to be added. This is perhaps a good place to comment on the use of Hall conductance as a probe of topological order: as demonstrated first by Kol and Read \cite{PhysRevB.48.8890}, in the absence of Galilean invariance -- for example, due to the presence of a lattice potential -- the connection between the filling and the Hall conductance is somewhat more subtle, particularly for more complicated fractions, such as $2/5$.\footnote{This was revisited in the context of fractional Chern insulators by Shankar and Murthy \cite{murthy2011composite,2012arXiv1207.2133M}.}
 
\begin{figure}\center
\includegraphics[width=0.5\columnwidth]{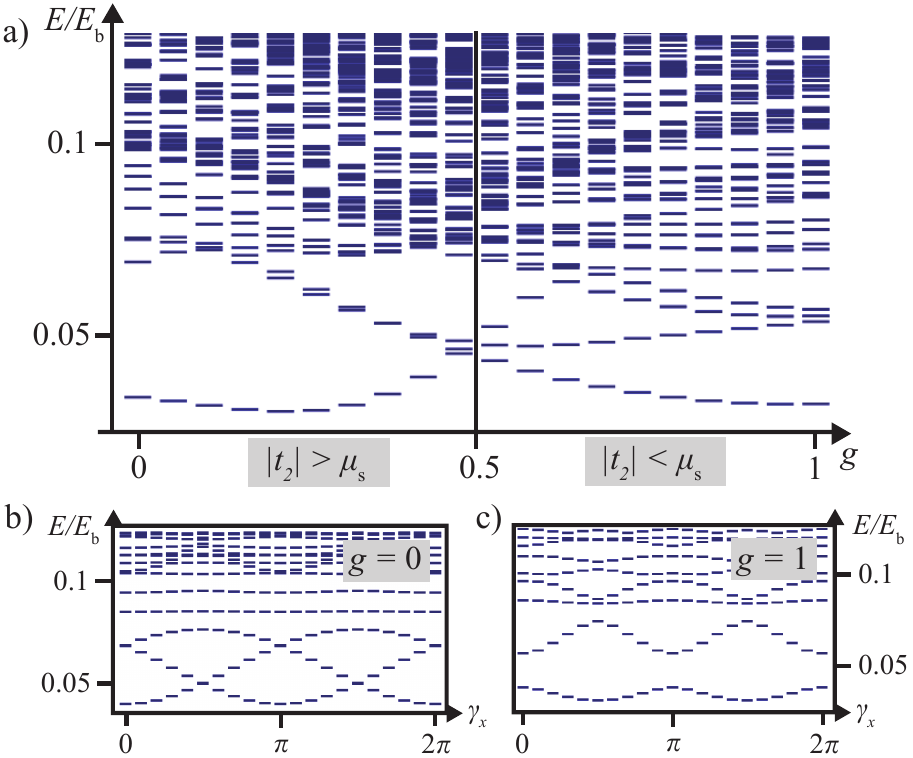}
\caption{{\bf Results from Ref.~\cite{Neupert:2011p1}}, illustrating (a) energy spectrum for $6$ electrons at $\nu=1/3$; (b)  spectral flow within the ground state multiplet in the fractional Chern insulator;  and (c)  the lack of such spectral flow  in a fractionally filled {\it trivial} insulator. \label{Neupert figure}}
\end{figure}

 \subsection{Quasihole Counting and Generalized Exclusion Statistics}
 A careful study of the quasihole excitations of  fractional Chern insulators was used by  Bernevig and Regnault~\cite{regnault2011fractional} to argue extremely convincingly for the existence of FQH states in the Chern band. After first establishing that the gap, ground state degeneracy and Hall conductance all remained robust with increasing system size, and that the ground state wave functions described uniform-density fluids (as opposed to modulated charge density waves), they proceeded to study the {\it excitations} of the putative FQH ground state. In additional to being fractionally charged, these are expected to exhibit fractional statistics both in their exchange with each other, as well as in the generalized `exclusion principle' that governs the counting of their low-energy many-body eigenstates \cite{PhysRevLett.67.937}. While fractional charge and exchange statistics are difficult to study in finite-size numerics, the counting of quasihole states can be directly verified.

 Assuming the system at commensurate filling to be in a particular FQH phase places strong constraints on the precise number of low-lying energy eigenvalues of the quasihole states in each generalized momentum sector. In the conventional FQHE, this counting can be rationalized in terms of a generalized exclusion principle \cite{PhysRevLett.67.937}. For example, working in the one-dimensional basis characteristic of the continuum Landau levels, the generalized exclusion principle for the $1/m$ Laughlin states forbids more than $1$ particle in $m$ adjacent orbitals. This leads to a precise prediction for the counting in each momentum sector (here by `momentum', we mean the appropriate conserved quantum number conserved by the symmetries of the chosen gauge-fixing.) Such a principle can be applied to each conserved momentum sector, and at a more coarse grained level the {\it total number} of low-energy  states in a finite system with a given, fixed number of quasiholes is also given by the exclusion principle. Note that these statements are true for the full spectrum only in the case of model Hamiltonians designed to render a particular FQH wave function and its quasiholes  exact eigenstates. For a generic Hamiltonian, they are expected to apply to the low-lying energy spectrum that is believed to capture the universal topological content of the phase, which is typically separated by a gap from spurious content at higher energies -- although such a gap may be small or absent for small system sizes or infelicitously chosen aspect ratios.

In order to rationalize the results for the Chern band, Bernevig and Regnault \cite{regnault2011fractional,PhysRevB.85.075128} gave an `unfolding' prescription to obtain a one-dimensional generalized momentum parameter to label the many-body eigenstates: for an $N_x\times N_y$ system, for instance, they labeled states in terms of $k \equiv k_x + N_x k_y$. With this procedure, it is possible to compare the counting of quasihole states both in total as well as within each momentum sector; such states were generated by diagonalizing the problem at a filling which deviates slightly from commensuration\footnote{The small system size means that this choice can be delicate; see \cite{regnault2011fractional} for details.}, corresponding to studying the system with a fixed number of quasiholes: for instance, diagonalizing the problem of $9$ electrons on a $5\times 6$ system introduces $3$ quasiholes of the $1/3$ state. The total  counting matched that predicted for the LLL quasihole states. Although the counting by momentum sector is somewhat more sensitive and occasionally exhibited a mismatch, in many cases this also conformed to the expectations from the LLL. This strongly bolstered the original numerical studies and provided the first evidence that the FCI excitations obeyed similar generalized exclusion statistics as their cousins in the LLL.

\begin{figure}\center
\includegraphics[width=0.8\columnwidth]{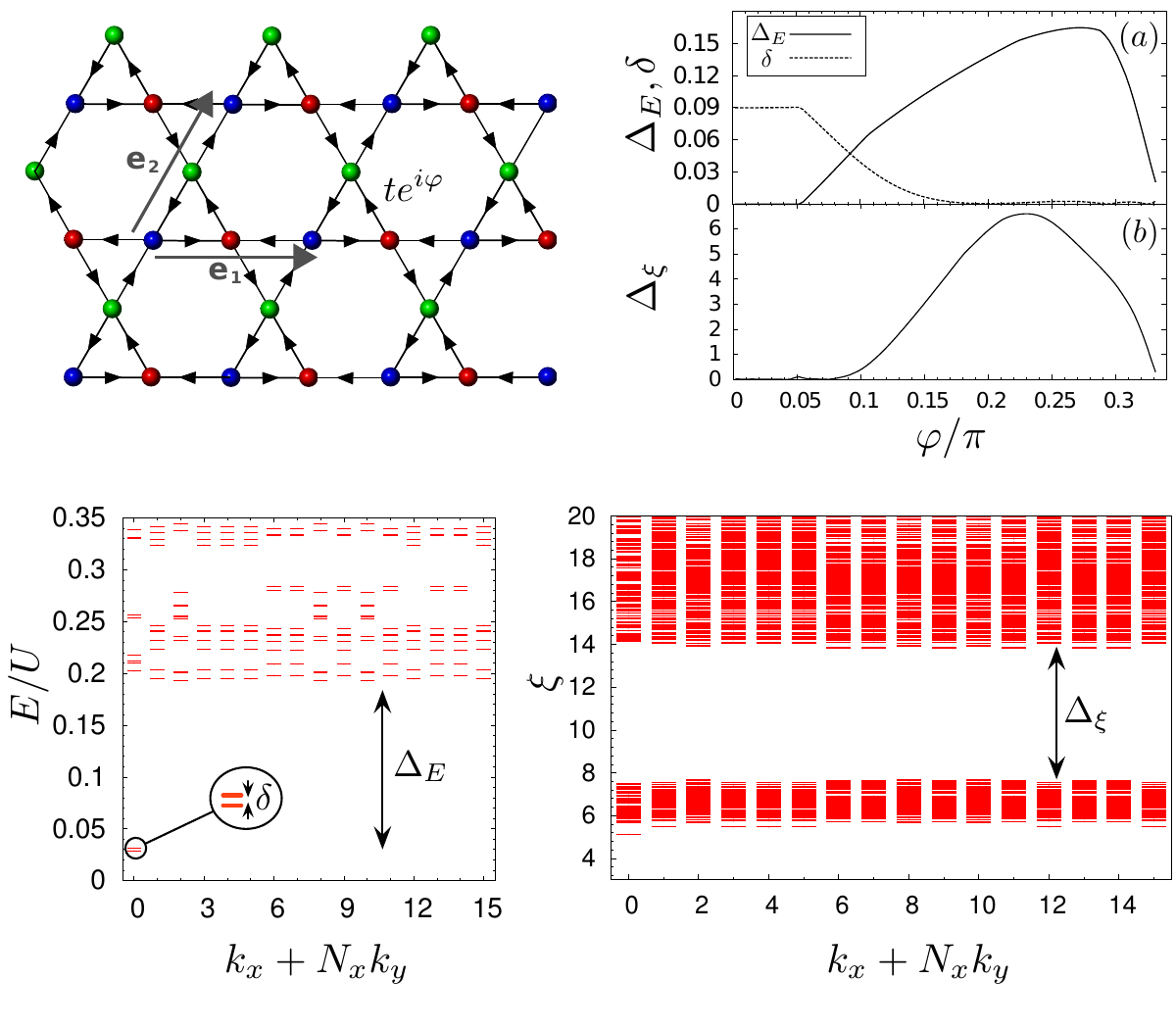}
\caption{{\bf Results from Refs.~\cite{regnault2011fractional,PhysRevB.85.075128} for $N=8$ bosons on an $N_x=N_y=4$ lattice ($\nu=1/2$ bosons).} Clockwise from top left: (i) kagom\'{e} lattice model, with tunable hopping phase $\varphi$; (ii)  (a) energy gap ($\Delta_E$) and finite-size splitting of degenerate eigenstates ($\delta$) and (b) entanglement gap ($\Delta_\xi$), as function of $\varphi$; (iii) particle entanglement spectrum keeping $N_A=4$ bosons, showing entanglement gap (counting of levels below gap match predictions from quasihole counting); and finally (iv) low-lying energy spectrum at $\nu=1/2$ showing ground state degeneracy (magnified, inset) and spectral gap.\label{fig:RBFig}}
\end{figure}

\subsection{Entanglement Spectrum}
A  powerful tool that has emerged in studying correlated topological phases, particularly in numerics, is the study of entanglement properties of the ground state \cite{PhysRevLett.96.110404,PhysRevLett.96.110405}. Various aspects of ground state entanglement have been discussed in the literature, and it is well beyond the scope of this review to give an exhaustive summary; here we will only provide a telegraphic account of the details needed to discuss the numerics. The study of the entanglement encoded in a ground state $\ket{\Psi}$ proceeds first by constructing the density matrix of the state, $\hat{\rho} = \ket{\Psi}\bra{\Psi}$. The next step is to divide (`cut') the Hilbert space into two disjoint subsystems $A$ and $B$: $\mathfrak{H} = \mathfrak{H}_A\otimes  \mathfrak{H}_B$, and construct the {\it reduced density matrix} of one subsystem, say $A$, by tracing over the degrees of freedom in the remainder:
\be
\hat{\rho}_A = \text{Tr}_{B} [\hat{\rho}]
\ee
The reduced density matrix captures the various properties of ground state entanglement. If we define\footnote{Note that $\hat{\rho}_A$ is positive definite so $\hat{H}_A$ is Hermitian.} an `entanglement Hamiltonian' $H_A$ we may write $\hat{\rho}_A =  e^{-\hat{H}_A}$,  and the set of eigenvalues of $\hat{H}_A$ constitute the {\it entanglement spectrum} of the ground state \cite{Li_08_010504}. Another commonly used measure is the {\it entanglement entropy} \cite{PhysRevLett.71.666}, which is the von Neumann ($S_{vN} = \text{Tr}_A [\hat{\rho}_A\ln \hat{\rho}_A]$) or Renyi ($S_{R}^{(n)} = \text{Tr}_A[ \hat{\rho}_A^n]$) entropy computed from the reduced density matrix.

Devoid of any other information, the entanglement spectrum does not offer significantly more information than the entanglement entropy; typically the best that can be hoped for is to understand the level statistics of the eigenvalue distribution. Where a knowledge of the entanglement spectrum comes into its own is when the system is endowed with an additional symmetry that can serve as an organizing principle for the entanglement eigenvalues; if the `cut' commutes with the symmetry, the entanglement spectrum is block diagonal in the corresponding symmetry sectors, and various additional properties can be extracted. For instance, in pioneering work Li and Haldane demonstrated that the entanglement spectrum of rotationally invariant quantum Hall states on the sphere could be organized using their conserved angular momentum. They matched the counting of the low-lying entanglement levels\footnote{Here and wherever we discuss entanglement spectra, by `low-lying' we mean states below the so-called `entanglement gap' that is generically observed to separate the universal content of the entanglement spectrum from spurious behavior at higher entanglement `energies'.} in each angular momentum sector with that of the chiral conformal field theory that describes gapless excitations at the physical edge of the corresponding FQH state. An enormous body of knowledge about the entanglement spectrum of FQH states in the LLL has  been built by analytical and numerical studies. 

Bernevig and Regnault \cite{regnault2011fractional} computed the ground-state {\it particle} entanglement spectrum -- which is obtained my making a `cut' in particle number space, i.e. producing a reduced density matrix by tracing out a certain number of particles rather than over a geometric region.\footnote{Note that in a periodic system  with topological order there is some ambiguity as to the treatment of degenerate eigenstates; Bernevig and Regnault chose to study a uniform incoherent superposition of the $q$ degenerate Laughlin states, but the phase dependence of the entanglement entropy of {\it coherent} superpositions can be used to extract additional topological properties\cite{PhysRevB.85.235151}.} The resulting entanglement spectrum was organized using a similar `unfolding' procedure as that used for the true spectrum. While the real-space entanglement spectrum is tied to the behavior of edge excitations of a FQH droplet, a particle space cut encodes the structure of quasihole states (at a filling corresponding to the number of particles left after the cut is made.) Thus, the particle entanglement spectrum should exhibit similar counting as the actual quasihole spectrum for the same number of quasiholes (see Fig.\ref{fig:RBFig}, bottom right, for an example).  Verification of this property, which has been demonstrated in detail in the LLL, for the fractionally filled Chern band constitutes another important indication that the two systems exhibit the same topological order. Furthermore, both the energy and entanglement spectra can be rationalized more precisely with an understanding of the emergent many-body symmetries (originally introduced by Haldane for the LLL) in the context of the Chern band~\cite{PhysRevB.85.075128}.
  
  \subsection{Extensions: Trial wave functions, Pseudopotentials, and Adiabatic Continuity}
  Since the early results that established the existence of FQH phases in Chern bands, there has been an explosion of numerical studies on fractional Chern insulators, an exhaustive list of which would be impossible to provide. Limitations of space prevent us from discussing several interesting extensions, such as to bosonic systems, the inclusion of disorder and numerical studies of bands of higher Chern number.\footnote{Although we briefly discuss a variational wavefunction analysis for phases in $C=2$ bands, below, and comment on some analytical results on lattice defects in such bands.} One development we {\it do} wish to briefly summarize,  especially as it connects naturally to the next section, is the study of trial wave functions and pseudopotential Hamiltonians. \footnote{We present this discussion here for reasons of organization; to the uninitiated, its clarity may be much improved by returning to it after reading the next section.} Here, Qi provided a route to constructing pseudopotential Hamiltonians and corresponding model wave functions in the Chern band that built on the understanding of those in the LLL via a mapping of Wannier states, discussed below. Subsequent detailed work extended this pseudopotential formalism \cite{2012arXiv1207.5587L}, studying the effects of inhomogeneous Berry curvature, as well as providing various prescriptions for improving the trial wave functions to account for the detailed structure of a Chern band.  Finally,  adiabatic continuity between Qi's model Hamiltonians and more realistic projected density-density interactions in a Chern band has been established numerically for certain FQH phases \cite{PhysRevB.87.035306}.

\section{Landau Levels and Chern Bands: A Critical Comparison}
As noted in the previous section, a variety of different models realize topological flat bands that host insulating phases at fractional fillings. On one level, this is extremely satisfying as it vindicates a certain prejudice: that since the IQHE has a lattice realization in a Chern band, so should the FQHE by analogy. This is bolstered by evidence such as the fractional Hall conductance, that are consistent with FQH physics. However the evidence we have reviewed so far has been predominantly numerical; it is clearly desirable to have a more analytical approach underpinning our understanding of the connection between Landau levels and Chern bands. Ideally this would achieve two goals:
\begin{enumerate}[i]
\item establish some notion of adiabatic continuity between the FCI ground state and model FQH wave functions in a Landau level.
\item provide a quantitative measure of the role of the lattice potential in the FCI. 
\end{enumerate} 

To see why such an analytical understanding is important, note that the restriction of the problem to the low-energy Hilbert space corresponding to the lowest Chern band follows from the first of our three criteria for a `good' Chern band. In simple terms, we have shown that it is reasonable to study the problem restricted to a space of states that have the same global topology (as quantified by their Chern number) as the orbitals in a single Landau level. If we can also demonstrate that a simple local Hamiltonian takes a similar form when written in the basis of Chern band eigenstates as in the Landau orbital basis, then we can argue using the basic notion of adiabatic continuity that the two problems represent essentially the same physics. There are two ways to accomplish this: the first, taken by Qi, is to construct a one-one-mapping between Landau gauge orbitals and (linear combinations of) Chern band eigenstates, and transcribe a Hamiltonian appropriate to a particular FQH phase from the LLL to the lattice Chern band. If this mapping is local, this would establish the existence of an FQH phase in the Chern band. An alternative approach, put forth by the present authors, is to begin with the same model of a projected density-density interaction in both cases, and demonstrate that the algebra of  projected density operators in the Chern band has the same features as the Girvin-Macdonald-Platzman algebra in the LLL. 
While the former approach permits a construction of trial wave functions, an advantage of the latter approach is that makes manifest the importance of the second criterion of Berry  flatness for a Chern band to reproduce LLL physics. A  more detailed study of the algebra also suggests the various constraints on the structure of the single-particle states in the Chern band that constitute the third and final criterion.

\subsection{Wannier Orbitals, wave functions, and Pseudopotential Hamiltonians}
The first concrete correspondence between Chern bands and Landau levels was proposed by Qi~\cite{Qi:2011p1}, who constructed lattice analogs of the Landau gauge single-particle eigenstates familiar from the continuum. Recall that the latter are given by
\be\label{eq:LLeigfunc}
\psi_{K_y}(\br) = \frac{1}{\pi^{1/4}\sqrt{\ell_B L_y}} e^{i K_y y}e^{-\frac{1}{2\ell_B^2}\left(x - K_y\ell_B^2\right)^2}
\ee
which are centered at $x = K_y \ell_B^2$, and $K_y = \frac{2 \pi n}{L_y}$ with $n$ an integer. Clearly, the center of this wavefunction shifts as $x_0 \rightarrow x_0 + \frac{2\pi}{L_y} \ell_B^2$ under transforming $K_y\rightarrow K_y +\frac{2\pi}{L}$. In the Chern band, Qi constructed Wannier states that have a one-dimensional momentum label with respect to which their centers evolve with a  similar `shift' property. We briefly summarize his Wannier construction for a Chern insulator placed on an infinite cylinder, periodic in $y$ and infinite in $x$. We will follow the notation of the previous section and consider a Chern band $\alpha$. It is always possible to perform a unitary (gauge) transformation on the single-particle states so that we fix one component of the Berry vector potential: $(\mathcal{A}_\alpha)_y= 0$. In this case the Wannier states in the Chern band are given by (suppressing the index $\alpha$)
\be
\ket{W(k_y, x)} &=& \frac{1}{L_x}\sum_{k_x} e^{-i \int_0^{k_x} \mathcal{A}_{x}(p_x, k_y)}
e^{-ik_x(x - \theta(k_y)/2\pi)}\ket{\bk}
\ee
where $\bk = (k_x, k_y)$,  $\theta(k_y) = \int_0^{2\pi}\mathcal{A}_x(p_x, k_y) dp_x$, and $x$ is an integer labeling the lattice sites. The phase factor involving $\theta(k_y)$ guarantees that the symmetry of the Bloch functions under $k_x\rightarrow k_x+2\pi$ is properly accounted for. Under a gauge transformation that multiplies the Bloch states by a phase factor $\ket{\bk} \rightarrow e^{i\phi(\bk)} \ket{\bk}$, the Wannier function is invariant upto a phase: $\ket{W_\alpha(k_y,x)} \rightarrow e^{i\phi(0,k_y)}\ket{W(k_y,x)}$. Most significantly, the center-of-mass position of the Wannier function $\ket{W(k_y,x)} $ is given by
\be
\langle\hat{x}\rangle = \bra{W(k_y,x)} \hat{x}\ket{W(k_y,x)}  =  x - \theta(k_y)/2\pi
\ee
-- in other words, $\theta(k_y)/2\pi$ is the shift of the Wannier function from the lattice site, often referred to as the charge polarization. However,  the integral of this quantity gives the Chern number, $C = -\frac{1}{2\pi} \int_0^{2\pi}\theta(k_y) dk_y$; thus, if $C\neq 0$, the Wannier function is not periodic in $k_y$. The center-of-mass position shifts from $x \rightarrow x+C$ when $k_y$ evolves from $0$ to $2\pi$, so that the Wannier functions satisfy twisted boundary conditions:
\be
\ket{W(k_y+2\pi, x)}  = \ket{W(k_y, x+C)}
\ee
We can define a generalized momentum coordinate $K_y = k_y +2\pi x$ for $0\leq k_y<2\pi$, in which case if we define   $\ket{W(K_y)} = \ket{W(k_y,x)}$, we have a continuously defined Wannier function for $K_y \in \mathbb{R}$. The center-of-mass position of the Wannier function now evolves continuously for $K_y$, which thus plays a role analogous to the conserved momentum of the Landau gauge eigenfunctions in the continuum. 
It may be verified that under the weak-field (Hofstadter) limit $\ell_B\gg1$, the Wannier functions $\langle \br |W(K_y)\rangle \rightarrow \psi_{K_y}(\br)$.

The next step is to recall that many model wave functions representing quantum Hall states in the LLL are exact ground states of local pseudopotential Hamiltonians \cite{Haldane:1983p1,Trugman:1983p1}. Working on the cylinder,  we can represent the model wave functions in the basis of occupation numbers of the single-particle states (\ref{eq:LLeigfunc}); using the correspondence   demonstrated above, this can be transcribed directly into a wavefunction on the lattice written in the basis of Wannier orbital occupation numbers. For instance, consider the LLL wavefunction for the $m^{\text{th}}$ Laughlin state on the cylinder
\be\label{eq:Laughlincylinder}
\Psi^{1/m}_{\text{LLL}}(\{z_i\}) &=& \langle\{z_i\} | \Psi^{1/m}_\text{LLL} \rangle
= \Omega \prod_{i<j}\left(e^{2\pi z_i/L_y} - e^{2\pi z_j/L_y}\right)^me^{-\sum_i x_i^2/2\ell_B^2}
\ee 
where $\Omega$ is a normalization factor and $z = x+i y$. Following Qi's prescription, the corresponding Chern band Laughlin wavefunction on the lattice takes the form
\be
|\Psi^{1/m}_{\text{CB}}\rangle = \sum_{\{n_i\}} \Phi(\{n_i\}) \prod_i\ket{W_{2\pi n_i/L_y}}
\ee
where the coefficients are obtained by expanding the wavefunction in the basis of Landau gauge eigenstates:
\be
\Phi(\{n_i\}) = \frac{1}{L_y^n} \int \prod_i d\br_i \psi^*_{2\pi n_i/L_y}(\br) \Psi^{1/m}_{\text{LLL}}(\{z_i\}).
\ee
The pseudopotential Hamiltonian that renders (\ref{eq:Laughlincylinder}) an exact ground state  can also be written in terms of the single-particle orbitals, and therefore readily transcribed to the Chern band. A final question is to determine if the resulting interacting lattice model is local. It is possible to demonstrate that the model Hamiltonian is indeed built from local terms on the underlying lattice. In this fashion, Qi produced a model Chern band  Hamiltonian and corresponding exact ground state that has the same topological order as a given quantum Hall wavefunction, thereby establishing the existence of the FQHE in a Chern band. In essence, the pseudopotential Hamiltonian  is a special point in the space of Chern band Hamiltonians where the FQHE is manifest. It is not clear {\it a priori} if this is related to the projected density-density interactions typically used in lattice models. The trial wave functions that result from this construction may be compared with the exact ground states of projected Hamiltonians; in doing so, some care is needed to gauge-fix the single-particle eigenstates in a manner that preserves the underlying lattice symmetries. For a detailed discussion, we refer the reader to the literature~\cite{PhysRevB.86.085129}.

\subsection{Projected Densities and $W_\infty$ Algebras}
An alternative perspective to Qi's Wannier orbital approach is to consider the { algebra} of densities projected into the Chern band which will allow a quantitative comparison between the LLL and the Chern band \cite{parameswaran2012fractional}. Recall that in the quantum Hall problem, the magnetic translation operators\footnote{We remind the reader that these are simply the projected densities stripped of a Gaussian factor $e^{-q^2\ell_b^2/4}$.} obey the so-called Girvin-Macdonald-Platzmann Algebra \cite{GMP:1986p1},
\be\label{eq:GMP}
[\overline{\rho}_{\bq_1},\overline{\rho}_{\bq_2}] =  2i \sin \left(\frac{\bq_1\wedge\bq_2\ell_B^2}{2}\right) \overline{\rho}_{\bq_1+\bq_2}
\ee

Consider the projected densities in a Chern band, given by (\ref{eq:CBprojdens});
at long wavelengths $qa\ll1$, we may expand $\sum_{b}  u^{\alpha *}_b\left(\bk+\frac{\bq}{2}\right)u^{\alpha}_{b}\left(\bk-\frac{\bq}{2}\right)
\approx 1 - i\bq \cdot \sum_b u_b^{\alpha *}(\bk) \frac{\nabla_\bk}{i} u^\alpha_b (\bk)$
$ \approx e^{i \int_{\bk-\bq/2}^{\bk+\bq/2} d\bk' \cdot\mathcal{A_\alpha}(\bk')}$, so that (here and below we suppress band indices)
\be\label{eq:approxparalleltransport}
\overline{\rho}_\bq \ket{\bk, \alpha} \approx e^{i \int_{\bk}^{\bk+\bq} d\bk' \cdot\mathcal{A_\alpha}(\bk')} \ket{\bk +\bq, \alpha} \ .
\ee
 In other words, for small $q$, $\overline{\rho}_\bq$ implements parallel transport described by the Berry connection 
  $\vec{\mathcal{A}}_\alpha(\bk)$. Either from this observation or via a gradient expansion, we may show that at long wavelengths, the commutator of projected density operators at different wavevectors is
\be\label{eq:Cherncomm1}
\left[ \overline{\rho}_{\bq_1},\overline{\rho}_{\bq_2}\right] &\approx& i \, \bq_1\wedge\bq_2 \sum_{\bk} \left[\mathcal{B}_\alpha(\bk)\sum_b   u^{\alpha*}_b\left(\bk_+\right)u^\alpha_b\left(\bk_-\right)
\times \gamma^\dagger_{\bk_+,\alpha}\gamma_{\bk_-,\alpha}\right]
\ee
where we define $\bk_{\pm} = \bk \pm \frac{\bq_1+\bq_2}{2}$. Finally, let us assume that the local Berry curvature $\mathcal{B}_\alpha(\bk)$ can be replaced by its average
\be\label{eq:Bbardef}
\overline{\mathcal{B}_\alpha} = \frac{\int_{BZ} d\bk\, \mathcal{B}_\alpha(\bk)}{\int_{BZ} d\bk } = \frac{2\pi C_\alpha}{{A}_{BZ}}
\ee
over the BZ; here ${A}_{BZ} = c_0^2/a^2$ is the area of the BZ, with $a$ the lattice spacing and $c_0$ a numerical constant depending on the unit cell symmetry.
This yields
\be\label{eq:Cherncomm2}
\left[ \overline{\rho}_{\bq_1},\overline{\rho}_{\bq_2}\right] \approx i \, \bq_1\wedge\bq_2  \overline{\mathcal{B}_\alpha}\,\overline{\rho}_{\bq_1+\bq_2}.
\ee
which is identical to the long-wavelength limit of the density algebra (\ref{eq:GMP}) for the LLL, with  $\overline{\mathcal{B}_\alpha}^{1/2} =\frac{\sqrt{2\pi C_\alpha}}{c_0} a$ playing the role of the magnetic length
$\ell_B$.
We may draw several inferences from Eqn.~(\ref{eq:Cherncomm2}), as detailed in previous work. We summarize them briefly here. First, we observe that a coarse grained, projected, position operator may be defined via $\br_{cg} \equiv x_{cg} \xhat + y_{cg}\yhat
={\displaystyle \lim_{\bq \rightarrow 0}} \frac{\nabla_q}{i} \overline{\rho}_{\bq}$. It follows from Eqn.~(\ref{eq:Cherncomm2}) that
\be
[x_{cg}, y_{cg}] =- i\overline{\mathcal{B}_\alpha}
\ee
which identifies the $\br_{cg}$ with the guiding center position operator in the LLL.

Note that without too much work we can  construct operators which obey the true GMP algebra while relaxing the long-wavelength approximation. For a system of $N$ unit cells there are $N$ points in the BZ. If $\mathcal{B}_\alpha(\bk)$ is truly constant we can define a set of $N$ parallel translation
operators $T_\bq$ for which (\ref{eq:approxparalleltransport}) holds exactly:
\be\label{eq:exactparalleltransport}
T_\bq \ket{\bk, \alpha} = e^{i \int_{\bk}^{\bk+\bq} d\bk' \cdot\mathcal{A_\alpha}(\bk')} \ket{\bk +\bq, \alpha} \ .
\ee
The algebra of the $T_\bq$ is thus exactly of the $\Winf$ form (\ref{eq:GMP}) without a long-wavelength restriction. We note that the $T_\bq$  are trivially isomorphic to magnetic
translation operators for a system with $N$ sites and flux $1/N$ per unit cell and it is straightforward to check that the states in the band form an $N$ dimensional irreducible
representation of their algebra \cite{Brown:1964p1}. From this perspective the idealization of a constant curvature Chern band hosts a $\Winf$  algebra whose long-wavelength generators coincide
with the physical density operators.

Finally, note that both the ``Hofstadter'' and ``Haldane'' problems give rise to (\ref{eq:Cherncomm2}), which unifies  earlier lattice FQHE studies \cite{Sorenson:2005p1,Moller:2009p2} with the ones considered here. In this sense, those earlier works can be seen as the first demonstration of lattice FQH phases, albeit those in which the Berry curvature is extrinsic and due to a magnetic field rather than an intrinsic property of the lattice model.
This last observation can be used to get nearly constant curvature bands by approaching the Landau level limit on the lattice, i.e. by picking flux
$1/m$ per plaquette and working the lowest subband at large $m$. While it is impossible to find a constant-curvature Chern band in  models of Chern insulators with $\mathcal{N}=2$ bands, it is possible to construct models with $\mathcal{N}>2$ which host a Chern band with nearly constant Berry curvature \cite{RahulUnpub,RegnaultBernevigUnpub}.

Readers familiar with the LLL problem will note that there the $\Winf$ algebra in a system with $N_\Phi$ states is generated by $N_{\Phi}^2$ density operators while in a Chern band for an $N\times N$ lattice system there are $N^4$ states
there are only $N^2$ densities (or $T_\bq$ if one wishes to work with a closed algebra). This distinction arises as the LLL is formally defined on a continuous space but
is without fundamental dynamical significance as the relevant momenta, $q \ell_B < 1$ are $O(N_\Phi)$ in the LLL as well. For example, it was shown in \cite{Boldyrev:2003p1} that
keeping only this set of momenta keeps the entire physics of the quantum Hall localization transition in the LLL. However this counting discrepancy does have the consequence
that the algebra of the densities themselves {\it must} close in the LLL at  {\it all} $\bq$ which is not the case in the Chern band. We will encounter this issue of operator counting again later in this review, when we discuss the Shankar-Murthy `Hamiltonian' approach to the fractional Chern insulator problem.

\subsection{Geometry of the Chern Band}
The algebraic approach to the Chern band discussed can be extended further to include effects of the band {\it geometry} in addition to the topology~\cite{Roy2012Geometry}. This allows further constraints to be placed on models in order that they stabilize fractionalized phases. While important to a complete understanding of the structure of Chern bands, this section uses somewhat more technical machinery than the remainder, and can be skipped on a first reading.  In the preceding section, we showed that to order $q^2$, and in the uniform Berry curvature approximation the algebra of projected densities obeys an analog of the GMP algebra of the LLL. A natural extension of this result is motivated by examining higher order terms in $q$ in the expansion of $
 [\bar{\rho}_{\bq_1},\bar{\rho}_{\bq_2}]$, and demanding that they in turn vanish. At order $q^3$, after some algebra we can demonstrate \cite{Roy2012Geometry} that this occurs if and only if a certain gauge-invariant quantity, known as the {\it Fubini-Study} (FS) {\it metric tensor},
$g^{\alpha}_{ij}(\bk)$ is a constant across the Brillouin zone. The FS metric is a rank two symmetric tensor, $g^{\alpha}_{ij}(\bk)$, which for band $\alpha$ has components~\cite{page1987geometrical,PhysRevLett.65.1697,kobayashi1969foundations,Pati1991105} 
\be
g^{\alpha}_{ij}(\bk)&=&\frac{1}{2}\sum_{a}\left[\left(\frac{\partial {u}^{\alpha*}_a}
{\partial k_i}
\frac{\partial{u}^{\alpha}_a}{\partial k_j} + 
\frac{\partial {u}^{\alpha*}_a}{\partial k_j}
\frac{\partial{u}^{\alpha}_a}{\partial k_i} \right)  - \sum_{b}\left(\frac{\partial {u}^{\alpha*}_a}{\partial k_j}u^{\alpha}_a{u}^{\alpha*}_b 
\frac{\partial {u}^{\alpha}_b}{\partial k_i} +  
\frac{\partial {u}^{\alpha*}_a}{\partial k_i}u^{\alpha}_a{u}^{\alpha*}_b \frac{\partial 
{u}^{\alpha}_b}{\partial k_j} \right) \right] 
\ee
where the $u^\alpha_a(\bk)$ are the single-particle eigenstates defined previously. 
It is worth pausing to expand a little on the meaning of this rather formidable expression. Recall that the Berry curvature of a band is a natural geometric quantity that emerges from considering the evolution of the eigenstates of the single-particle Hamiltonian $h_{ab}(\bk)$ as the parameter $\bk$ evolves (adiabatically) across the BZ. Just as we defined the projector onto band $\alpha$ we can define the {\it orthogonal} projector $\mathcal{Q}_\alpha(\bk) =  \mathbf{1} - \sum_{\beta\neq \alpha}\ket{\bk,\beta}\bra{\bk,\beta}$.
It is convenient to introduce the (complex) tensor
\be
\mathcal{R}^{\alpha}_{ij}(\bk) \equiv  \left[\partial_{k_i} \bra{\bk,\alpha}\right]\mathcal{Q}_\alpha(\bk)\left[\partial_{k_j}\ket{\bk,\alpha}\right].
\ee
By rewriting the eigenstates in terms of the $u^\alpha_a(\bk)$ it is readily verified that the Berry curvature $\mathcal{B}_\alpha(\bk) = -i\epsilon^{ij}\mathcal{R}^{\alpha}_{ij}(\bk) =    -2 \,\text{Im}\left[\mathcal{R}^{\alpha}_{xy}(\bk)\right]$. \footnote{Note that the curvature also has indices, but in $d=2$ there is only one nonzero component so we can suppress them.} A little more work shows that the FS metric is simply the corresponding real part: $g^{\alpha}_{ij}(\bk) =\text{Re}\left[ \mathcal{R}^\alpha_{ij}(\bk)\right]$.\footnote{Observe that owing to the equivalence between the position operator $\hat{x}_i$ and the $\bk$-space derivative $-i\partial/\partial k_i$ it is possible to rewrite  $\mathcal{R}^\alpha_{ij} = \bra{\bk,\alpha}\hat{x}_i \mathcal{Q}(\bk) \hat{x}_j \ket{\bk,\alpha}$; in this form we see that the non-vanishing of the Berry curvature roughly corresponds to the non-commutativity of the $x$- and $y$-components of the position operator.} Note that the gauge-invariance of both the metric and the curvature are manifest in the above form.

The importance of the FS metric is that it introduces a natural notion of `distance' in the single-particle Hilbert space that can induced from the $\bk$-space evolution of the single-particle eigenstates.\footnote{This is intimately connected to the geometry of K\"ahler manifolds as well as to the geometric theory of the insulating state, but unfortunately we cannot discuss these beautiful subjects here \cite{2011EPJB...79..121R}.} Furthermore, since the FS metric and the Berry curvature are the real and imaginary parts of the {\it same} complex tensor $\mathcal{R}^\alpha(\bk)$, it is also possible to show that at any $\bk$ in the BZ, 
 the trace of the FS metric is bounded from below by the magnitude of the Berry curvature at $\bk$, 
\be  \text{tr}(g^{\alpha}(\bk)) \ge | \mathcal{B}_{\alpha}(\bk)| \label{eq:metric-berry-inequality}.
\ee
This geometrical constraint applies to {\it any} insulator. In the remainder of this section, we will use this in conjunction with the existence of a nonzero Chern number, to sharpen our understanding of the special properties of the restricted Hilbert space of a Chern band.
 
 We were led to consider the metric tensor by studying higher order terms in the expansion of the algebra of projected density, and we identified its uniformity as a criterion for this to match the $W_{\infty}$ algebra of the LLL to order $q^3$. Thus, we have found an additional criterion for identifying ``good" band structures
 from the point of view of hosting interacting topological
 phases. A natural question to ask is whether one obtains an infinite set of such constraints, order-by-order in $q$; it would be somewhat deflating if this were the case. Remarkably, we will demonstrate that when the band structure satisfies {\it one}
 additional constraint, the Chern band projected densities satisfy the $W_{\infty}$ algebra of projected LLL densities at {\it all} orders in $q$.
 
 To this end, it is useful to consider the transformation properties of $\mathcal{R}^\alpha_{ij}$ under unimodular transformations of the coordinates, i.e. those of the form $x_a \rightarrow x_a' = \Lambda_{ab} x_b$ with $\det\Lambda=1$. The FS metric transforms in the usual way, $(g^\alpha_{ab})'(\bk) =\Lambda_{ac} \Lambda_{bd} g^\alpha_{cd}(\bk)$, while the Berry curvature is invariant.  Moving to `primed' coordinates in which $g^\alpha(\bk)$ is diagonal,\footnote{For $d=2$ in this basis simple algebra shows that the square of the trace is four times the determinant.} using the invariance of the determinant under unitary transformations and applying the inequality (\ref{eq:metric-berry-inequality}) we obtain
 \be
2\sqrt{\det g^\alpha(\bk)} = 2\sqrt{\det(g^\alpha(\bk))'}   = \text{tr}(g^\alpha_{ij}(\bk))' \geq |[\mathcal{B}^\alpha(\bk)]'| = |\mathcal{B}^\alpha(\bk)|
 \ee
 Since this is true across the BZ, we can integrate this to find
 \be
 \int_{BZ} d^2k\, \det g^\alpha(\bk) \geq \frac{1}{4} \int_{BZ} d^2k\, |\mathcal{B}_\alpha(\bk)|^2 \geq  \frac{1}{4} A_{BZ}\bar{\mathcal{B}}^2_\alpha = \frac{\pi^2 C_\alpha^2}{A_{BZ}}\label{eq:final-ineq}
 \ee
 where we use the fact that the square-averaged Berry curvature is bounded from below by the square of the Chern number per unit area of the BZ. Eq. (\ref{eq:final-ineq}) is the key  relation between the Chern band geometry to its nontrivial topology: namely,  the integral of the determinant of the Fubini-Study metric is bounded from below by a number proportional to the square of
the Chern number of the band.

We now consider the case when (i) the inequality~(\ref{eq:final-ineq})
is saturated and (ii) the FS metric is uniform in the BZ. It is easily seen that these conditions amount to requiring, in addition to the uniformity of the Berry curvature obtained previously, the single additional constraint that 
$\det(g^{\alpha}(\bk)) = \frac{1}{4}{|\mathcal{B}_{\alpha}(\bk)|^2}$ at all 
$\bk$ in the BZ. In this limit and in the `primed' coordinate system used previously, in which the FS metric is diagonal, we have $[g_{ij}^\alpha(\bk)]' = \frac{\bar{\mathcal{B}}_\alpha}{2} \delta_{ij}$. Writing the density operators in the rotated coordinate system, it is straightforward to demonstrate that they satisfy a generalized, metric-dependent version of the $W_{\infty}$ algebra:
\be\label{eq:genWinf}
[\bar{\rho}_{\bq_1},\bar{\rho}_{\bq_2}]= 2i \sin\left(\frac{\bq_1 \wedge \bq_2 \bar{\mathcal{B}}_{\alpha}}{2}\right)e^{(q_1)_l g^{\alpha}_{lm}(q_2)_m}\bar{\rho}_{\bq_1 +
  \bq_2} = 2i \sin\left(\frac{\bq_1 \wedge \bq_2 \bar{\mathcal{B}}_{\alpha}}{2}\right)e^{\frac{\bar{\mathcal{B}}_\alpha}{2}(\bq_1+\bq_2)^2}\bar{\rho}_{\bq_1 +
  \bq_2} 
\ee
There are three important aspects of this result worth reflecting on. 
First, note that Berry curvature and the FS metric both
appear in the form of the $W_{\infty}$ algebra obtained in (\ref{eq:genWinf}), which thus
algebra also applies to bands which have a higher Chern number
and therefore differ fundamentally from Landau levels
with $C=1$. Second, observe that the
conditions under which we get a closed algebra of the projected
density operators can be stated purely in terms of the FS
metric. Third, we observe that for the $C=1$ case, (\ref{eq:genWinf}) looks remarkably similar to the algebra of projected densities that obtains in a  Landau level, with the combination $e^{(q_1)_l g^{\alpha}_{lm}(q_2)_m}$ playing a role analogous to the LL-dependent form factor that relates projected densities to the magnetic translation operators (which are identical in every Landau level) for the continuum FQHE. A heuristic interpretation of the additional constraints is therefore that they quantify not simply how close Chern band projected densities are to realizing the continuum Landau level algebra, but in addition how close they are to realizing the {\it lowest} Landau level.

For a system where the ideal conditions under which this
algebra is obtained do not hold, the degree of deviation from
these conditions provides an additional criterion to quantify how favorable a Chern band is for hosting
FQH-like physics. Conversely, if one finds fractional
topological phases in systems where the deviations from these
conditions is considerable, one could argue that the physics of
those systems is new and distinct from the conventional
fractional quantum Hall effect. The effects of disorder also enter the Hamiltonian through
terms that involve the projected density operator. This
suggests that the effects of disorder in the Chern band are
likely to be the same as in the LLL when the conditions stated
above for the FS metric are satisfied. Extensions of the algebraic approach to higher dimensions and to  time-reversal symmetric $\mathbb{Z}_2$ topological insulators have also been discussed in the literature \cite{PhysRevB.86.035125,PhysRevB.86.195125,PhysRevB.86.241104}.

\subsection{Operator Counting, Smoothed Densities, and the Shankar-Murthy Approach} 

We now return to the question of operator counting that we discussed earlier. As motivation, we first observe that in cases where the deviation of the Berry curvature from its average value is bounded, $|\mathcal{B}_\alpha(\bk) -\overline{\mathcal{B}_\alpha}| < |\overline{\mathcal{B}_\alpha}| -\epsilon$, we may define a  `smoothed' density operator which may be regarded as the projection of an operator $\rho^s(\br)$ local in position space; for $qa\ll1$  this gives a modified form of (\ref{eq:approxparalleltransport})
\be\label{eq:smootheddensities}
\overline{\rho}^s_\bq \ket{\bk,\alpha} = \frac{\overline{\mathcal{B}_\alpha}}{{\mathcal{B}_\alpha(\bk)}}e^{i \int_{\bk}^{\bk+\bq} d\bk' \cdot\mathcal{A_\alpha}(\bk')} \ket{\bk +\bq, \alpha}.
\ee
At long wavelengths, the algebra  of smoothed densities closes, and in this limit (\ref{eq:Cherncomm2}) is an {\it exact} equality when $\overline{\rho}_\bq$ is replaced by $\overline{\rho}^s_\bq$. This provides a clue that there are observables in the Chern band that indeed satisfy a well-behaved GMP algebra, related to the projected density operators in some nontrivial fashion.

Building on the different operator counting  in the LLL versus the Chern band, Shankar and Murthy were able to take a slightly different perspective on the operator algebra approach to FCIs. Their approach (described in Ref. \cite{2012arXiv1207.2133M,murthy2011composite}) consists of three separate steps:
\begin{enumerate}[(i)]
\item They construct a set of one-body operators $\rho_\text{GMP}(\bq+\mathbf{G})$ which act only within a Chern band, and satisfy the full GMP algebra (\ref{eq:GMP}). Here, $\bq$ is a crystal momentum (restricted to lie in the first BZ) while $\mathbf{G}$ is a reciprocal lattice vector. For an $N\times N$ lattice, the number of independent crystal momenta is $N^2$, and Shankar and Murthy demonstrated that the `umklapp' operators $\rho_{\text{GMP}}(\bq+\mathbf{G})$ are independent only for $N^2$ choices of reciprocal lattice vector $\mathbf{G}$. Thus, together there are $N^4$ independent GMP operators ${\rho}_\text{GMP}(\bq +\mathbf{G})$  which thus form a complete basis in terms of which one may express any of the $N^4$ one-body operators in a Chern band.
\item They then demonstrate that the projected density operators in the Chern band may be rewritten in terms of the $\rho_\text{GMP}(\bq +\mathbf{G})$:
\be
\overline{\rho}_\bq = \sum_{\mathbf{G}} c(\mathbf{G},\bq) \rho_\text{GMP}(\bq +\mathbf{G})
\ee
where the expansion is over reciprocal lattice vectors and the coefficients are determined from data on the lattice realization of the original Chern band. Thus the projected densities are a specific linear combination of operators satisfying the GMP algebra -- which explains why the $\overline{\rho}_\bq$ themselves only satisfy (\ref{eq:GMP}) in an idealized limit. They also noted a distinction between the projected density and the exponential of the guiding center density in a Chern band (in the LLL these are the same, up to a normalization depending only on $q$). Recall that an analogous situation arose in our discussion of the Chern band geometry  in the preceding section.
\item The preceding observations allow them to express a Hamiltonian projected to the Chern band in terms of operators that obey the GMP algebra. Note that since one may naturally formulate the full LLL problem in terms of the GMP operators, this suggests another route to constructing model Hamiltonians for a given FQH phase. Furthermore, Shankar and Murthy demonstrate that the `Hamiltonian theory' that they pioneered for the FQH -- in which Chern-Simons flux attachment is placed on an operator footing -- can be ported to the Chern band by working with the $\rho_\text{GMP}(\bq)$. This in turn permits the computation of gaps and response functions within a Hartree-Fock mean field theory.
\end{enumerate}
For reasons of brevity we cannot give a detailed elucidation of these points, and encourage the reader to study the original articles \cite{2012arXiv1207.2133M,murthy2011composite} for a pedagogical discussion of the Hamiltonian theory of fractionally filled Chern bands.

\subsection{Adiabatic Continuity and Establishing Topological Order}
\begin{figure}\center
\includegraphics[width=0.5\columnwidth]{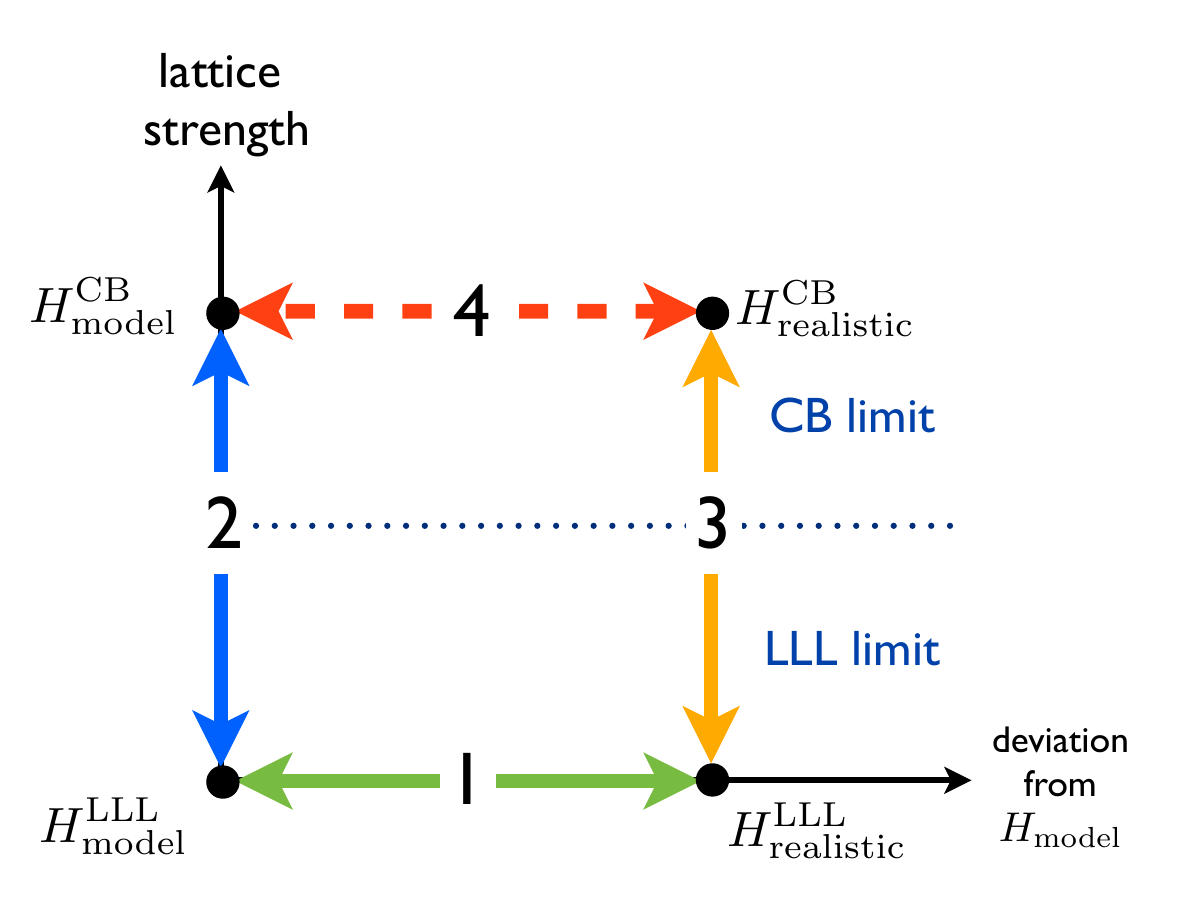}
\caption{\label{fig:adiabatic}{\bf Mappings within and between the LLL and the Chern band.}
{\bf Path 1: }Within the LLL, there exist special `model Hamiltonians' designed to a render a certain FQH state an exact ground state; a wealth of numerical evidence suggests that in many cases these are adiabatically connected to the ground states of more realistic Hamiltonians, for instance those obtained by projecting a density-density interaction to the LLL. {\bf Path 2: }Qi and Shankar-Murthy both construct mappings from the LLL to the Chern band that allow them to construct model Hamiltonians in the Chern band, with the {\it same} topological order in the LLL. {\bf Path 3: }The present authors instead considered realistic density-density interactions projected to the Chern band, and gave conditions for this to be adiabatically connected to the FQH ground state of a similarly realistic Hamiltonian in the LLL . {\bf Path 4: } Recent numerical studies suggest that the model Hamiltonians are adiabatically connected to these more realistic projected density-type interactions.}
\end{figure}
We have provided three different analytical perspectives on the correspondence between Chern bands and the (lowest) Landau level. We now place these in context by describing how they are related to the problem of establishing the topological order of the ground state of an interacting, fractionally filled Chern band. For this, it is useful to recall that at a filling for which a quantum Hall state is known to exist, there are typically two points in the space of all LLL-projected Hamiltonians that are of especial interest. The first is the model interaction, which we  denote $H_\text{model}^\text{LLL}$, that renders a certain FQH wavefunction an exact gapped ground state \footnote{Note that not all FQH states have parent Hamiltonians of this form;  here we shall restrict ourselves to those that do.}. The second is the one that is reached by projecting a realistic electronic interaction Hamiltonian into the LLL, which we shall call $H_\text{realistic}^\text{LLL}$. \footnote{Of course, there may be other details, such as the range of the interaction, etc., but we shall ignore this subtlety and discuss all `realistic' Hamiltonians in a unified fashion.} There is much numerical evidence establishing that in many cases the ground states of $H_{\text{realistic}}^\text{LLL}$ are connected adiabatically to those of $H_\text{model}^\text{LLL}$, thus establishing that for experimentally relevant and reasonable interactions the ground state has the same topological order as that encoded by the model system.

What about the Chern band? Here, we must first produce model Hamiltonians $H_\text{model}^\text{CB}$ on the lattice, since those constructed in the continuum rely intimately on the analytic structure of the LLL wave functions and cannot be ported directly to a Chern band. Here, Qi's prescription for mapping Landau gauge eigenstates from the LLL to the Chern band provides one route. Another is that of Shankar and Murthy; by suitably picking a pseudopotential interaction in the LLL and replacing the corresponding projected density operators by the $\rho_\text{GMP}$s, they can construct model Hamiltonians with similar topological properties in the Chern band. However, both these approaches can lead to Hamiltonians which may look somewhat unnatural on the lattice -- Qi's construction because the Wannier states may not be related by a simple transformation to the densities,\footnote{Recall that in a Chern band, Wannier orbitals can be chosen to be eigenstates of only one position coordinate.} they are not and Shankar and Murthy's approach because of the nontrivial relation between the $\rho_\text{GMP}$s and the projected density. 

A different approach, suggested by the present authors, attempts to simply study the properties of $H^\text{CB}_\text{realistic}$, defined by projecting a density-density interaction of some fixed range to the Chern band. Here, the mismatch between the CB and the LLL is encoded in the structure of the algebra of projected densities: as we have emphasized, they obey the GMP algebra only in certain idealized limits. Nevertheless, this suggests that when conditions are such that this idealization is not too far-fetched, one can argue that the ground states of $H^\text{LLL}_\text{realistic}$ and $H^\text{CB}_\text{realistic}$ will be in the same phase, and draw on the continuity between $H^\text{LLL}_\text{realistic}$ and $H^\text{LLL}_\text{model}$ in the continuum Landau problem to establish the topological order of the Chern band ground state for realistic interactions. 
Alternatively, one may view the idealization as placing constraints on various parameters, such as Berry curvature fluctuations, for there to exist an adiabatic path in the Chern band connecting $H^\text{CB}_\text{realistic}$ to $H^\text{CB}_\text{model}$. We note that numerical studies have demonstrated that when these constraints are violated, the stability of the Chern band ground state is adversely affected; also, the adiabatic continuity between Qi's model Hamiltonians, and more realistic interactions, has recently been verified \cite{PhysRevB.87.035306}.

 \section{Parton Constructions and Phase Transitions}
Thus far, we have focused on matching the physics of a Chern band with that of the lowest Landau level, and in establishing the existence of lattice analogs of familiar FQH states. Our survey of the existing analytical and numerical evidence should at this point have convinced the reader that at least some of the phases seen in the lowest Landau level survive to the lattice limit. In this section and the next, we change perspective slightly and ask whether the presence of the lattice can lead to additional physics beyond that familiar from the Landau level. A natural method to analyze such question is to use the `parton' construction or projective construction, a mean field theory that has had considerable success in analyzing both continuum FQH phases as well as lattice spin liquid models. As has been our theme overall, we will attempt only to give a flavor of this vast area, and  identify a few interesting results that have emerged from such studies.

\subsection{Partons and Projective Symmetry Groups}
We will give a very simplified account of a parton construction~\cite{PhysRevB.85.165134} of the $\nu=1/m$ Laughlin state, first in the continuum and then its lattice analog. 
In the continuum, the basic observation is that as a consequence of the dependence of the quantum of flux on the electric charge, a Landau level that is at $1/m$ filling for particles of charge $e$, is fully filled for particles of charge $e/m$. In other words, an identical { number} of $e/m$-charged particles would have a unique, Slater determinant ground state. In order to put this observation to use, we divide each electron into $m$ fermionic `partons', each of charge $e/m$, distinguished by a flavor index, i.e.  we write the electron operator as
\be
c(\br) =  \prod_{\alpha=1}^m f_\alpha(\br) \label{eq:partondecLLL}
\ee
We can now write mean-field states for each parton flavor independently. However, in splitting apart the electron and treating the resulting partons independently, we artificially enlarged the Hilbert space. To amend this, the final step of the construction is to project the mean-field wavefunction back to the physical subspace, by enforcing the requirement that any valid many body wavefunction for the physical electrons have $m$ partons at the same position coordinate. In other words, we write for the electron wavefunction
 \be
 \Psi^{1/m}_e\left(\left\{\br_i\right\}\right) = \bra{0} \prod_{i=1}^N \prod_{\alpha=1}^m f_\alpha(\br_i) \ket{\text{MF}}.\label{eq:partonglue}
 \ee
 Implementing this procedure for the LLL in the disc geometry, and using the fact that the charge $e/3$ partons fill Landau levels as discussed above, we find that 
 \be
 \Psi^{1/m}_e\left(\left\{\br_i\right\}\right) =\left[\Psi^{\nu=1}_\text{parton}\left(\left\{\br_i\right\}\right)\right]^m = \prod_{i<j} (z_i-z_j)^{m} e^{-\sum_i |z_i|^2/4\ell_B^2}
 \ee
 where $z\equiv x+iy$, ${\ell_B} =  \sqrt{\hbar/e|B|}$ is the electron magnetic length, and we used the fact that a filled LLL of charge $e/3$ partons has the Slater determinant wavefunction $\Psi^{\nu=1}_\text{parton}\left(\left\{\br_i\right\}\right) = \prod_{i<j} (z_i-z_j) e^{-\sum_{i} |z_i|^2/\tilde{\ell}_B^2}$, where $\tilde{\ell}_B = \sqrt{m}\ell_B$ is the magnetic length of the partons.
It is useful also to consider a field-theoretic formulation of the parton construction. The LLL parton decomposition (\ref{eq:partondecLLL}) of the electron has a $SU(m)$ gauge symmetry: rotations in the internal space that mixes the partons yield the same electron operator, and such rotations can be performed independently for each $\br$. Essentially, within the field theory the fluctuating gauge field emerges as a way of enforcing the local constraint that glues together partons to form physical electrons. The emergent gauge symmetry is reflected by the symmetries of the mean-field Hamiltonian that describes the parton, which is simply three copies of the kinetic energy of a charge $e/m$ particle in an external field; since the partons are gapped (fill a Landau level), they may be integrated out, yielding an $SU(m)_1$ Chern-Simons theory, that yields the correct ground-state degeneracy on the torus, and other topological properties. Note that the assumption of full $SU(m)$ symmetry is not strictly necessary; it is possible also to consider terms in the mean-field Hamiltonian that mix partons and break the symmetry down to the discrete $Z_m$ subgroup, but in the restricted Hilbert space of the LLL, both these symmetries give rise to the same electronic wavefunction after projection. 

How does the parton approach fare on the lattice? Our discussion will follow that of Lu and Ran ~\cite{PhysRevB.85.165134}, who first applied the parton construction to Chern insulators, but once again we only give the broad outlines. We begin by performing a similar decomposition (\ref{eq:partondecLLL}) of the electron operator on each lattice site into three partons, assuming for simplicity that each site has only one orbital (but of course, multiple sites within each unit cell.) A generic mean-field Hamiltonians for the partons takes the form
\be
H^\text{MF} =  \sum_{\br,\br'}\sum_{\alpha,\beta=1}^m f^\dagger_{\alpha}(\br) M_{\alpha\beta} ({\br,\br'}) f_\beta(\br) \label{eq:MFpartCB}
\ee
where $M({\br,\br'})$ is an $m\times m$ matrix. Under a local $SU(m)$ gauge rotation that transforms partons via $f_\alpha(\br) \rightarrow G_{\alpha\beta}(\br) f_\beta(\br)$ $M$ transforms as $M(\br,\br') \rightarrow G(\br) M(\br,\br') G(\br')$, with multiplication in the parton indices implied in the latter expression. Once again, the final electronic wavefunction is obtained by diagonalizing the quadratic parton Hamiltonian (\ref{eq:MFpartCB}) and  projecting to the physical degrees of freedom by imposing the constraint that $m$ partons must always appear togther, once again using (\ref{eq:partonglue}). We turn now to the delicate question of how to choose $H^\text{MF}$ to describe a fractional Chern insulator. 

As motivation, it is useful to note that for a $\nu = 1/m$ Chern band, there are is on average $1$ electron per $m$ unit cells; thus, a mean-field {\it electronic} wavefunction will generically lead to a metallic state. Ideally we would like each of the $m$ partons to fill a band with Chern number $1$; this would replicate our construction in the continuum. However, in the latter case, there was a simple expedient, charge fractionalization, which allowed us to change the effective filling via a redefinition of the quantum of flux for the partons. Here, the presence of the lattice makes this more subtle: essentially, the increase in the quantum of flux for the partons implies that, although the {\it electron} band structure respects all the lattice symmetries, the {\it partons} have a fraction of a flux in each plaquette of the original lattice. How are we to construct a symmetric wavefunction?

The key lies in the notion of the {\it projective symmetry group} familiar from slave-particle theories of spin liquids \cite{PhysRevB.65.165113}. In essence, the {\it physical} degrees of freedom are the electrons, so that a broken symmetry for the partons that is invisible to observables written in terms of electron operators is irrelevant. For instance, two mean-field solutions that differ by gauge transformations are equivalent as they yield identical electronic wave functions after projection; similarly, mean-field states may break lattice symmetries, and yet give rise to symmetric electronic wave functions. In mathematical terms, the various mean-field ansatze form projective representations of the symmetry group, and can be classified using well-developed technology. 

Following this general line of reasoning, the path to constructing symmetric wave functions at $\nu=1/m$ is as follows: we obtain a mean-field parton band structure by enlarging the unit cell on the original lattice, for instance by sticking $2\pi/m$ flux through each original unit cell. Since there was originally one electron per $m$ unit cells of the original lattice, but in the mean-field description each electron has been decomposed into $m$ partons, and the new unit  cell consists of $m$ of the original cells, it follows that the mean-field state we construct in this fashion has $m$ partons filling the $m$ lowest bands, and the resulting parton state is gapped. Note that counting in terms of the electrons, this corresponds to one electron per (enlarged) mean-field unit cell so that we expect the gap to be robust under the projection back to electronic degrees of freedom. Since each parton has charge $e^* = e/m$, and there are $m$ filled $C=1$ bands, we obtain the desired Hall conductance
\be
\sigma_{xy} = {m} \frac{(e^*)^2}{h} =  \frac{1}{m} \frac{e^2}{h}
\ee
 Although we have assumed each of the $m$ parton bands has $C=1$, note that different choices can give states where the filling and the Hall conductance do not bear a simple relation to each other \cite{PhysRevB.48.8890,2012arXiv1207.2133M,murthy2011composite}. It is also instructive to view the construction in terms of an explicit wavefunction. The electronic wavefunction is the projection of the Slater determinant for $m$ filled parton bands:
 \be
 \Psi^{1/m}_{e; Z_m}\left(\left\{\br_i\right\}\right) = \mathcal{P}\det \mathcal{W};\,\,\,\,\,\,\,\,\, \mathcal{W}_{(\alpha-1)N+i, (n-1)N+j}=  u_{\bk_j}^{(n)} (\br_i^\alpha) \label{eq:partonZm}
 \ee
 where $\mathcal{P}$ implements the constraint,  and $\mathcal{W}$ is an $mN\times mN$ matrix, with $i, j =1, 2,\ldots, N$ and $\alpha, n =1,2,\ldots,m$. Here, $u_{\bk_j}^{(n)}(\br_i^\alpha)$, is a single-particle momentum-$\bk_j$ state in which parton $\alpha$ is in the $n^\text{th}$ band (counting from below) obtained by diagonalizing the mean-field parton band structure. 
When the mean-field band structure has full $SU(m)$ symmetry, $M_{\alpha\beta}(\br,\br') =  \tilde{M}(\br,\br')\delta_{\alpha\beta}$, the lowest $m$ bands of the parton band structure are identical and we have $u_{\bk_j}^{(n)}(\br_i^\alpha) = u_{\bk_j}^{(n)}(\br_i^{(n)})\delta_{n,\alpha}$. The general form (\ref{eq:partonZm})  then factors into the $m$-th power of a Slater determinant:
\be
 \Psi^{1/m}_{e; SU(m)} =  \left[\det[ u_{\bk_j}(\br_i)]\right]^m \label{eq:partonSUm}.
\ee
The fully $SU(m)$ symmetric case is thus analogous to the case of the LLL. However, in general,  tunneling terms between the partons are present and rule out such a factoring, restricting us to $Z_m$ parton wave functions of the form (\ref{eq:partonZm}). This distinction between the $Z_m$ and $SU(m)$ wave functions is one way in which the parton construction on the lattice differs from that in the continuum LLL.

From the preceding discussion it should be clear that there is quite a bit of freedom in implementing this procedure; ideally we would like to be able to exhaustively list all such (mean-field) wave functions. Here, the full technology of projective symmetry group (PSG) classification is required, and we refer the reader to~\cite{PhysRevB.85.165134}, as well as related work~\cite{PhysRevB.85.125105,PhysRevB.85.165134,PhysRevB.86.075136} for details. We content ourselves with noting that the presence of the lattice allows the possibility of `symmetry-enriched' topological phases, whose low-energy descriptions have similar topological structure but realize lattice symmetry in distinct ways. 

\subsection{Bandwidth-Tuned Phase Transitions}
A fractionally filled  Landau level in the continuum limit has no unique ground state in the absence of interactions. In contrast, a Chern band (alternatively, a Landau level with a strong periodic potential) typically has a nonzero dispersion, characterized by its bandwidth $t$ (while perfect flat bands have $t=0$,  these are fine-tuned.) This weak dispersion can select a ground state in the $U/t \rightarrow 0$ limit, where $U$ is as usual the interaction scale; at partial filling, this ground state is expected to be compressible. \footnote{Note that this is the expectation whether the particles are bosons (a superfluid state) or fermions (a filled Fermi sea).} As $U/t$ is increased, eventually we enter the flat band limit where the kinetic energy can be ignored; we assume the system forms an incompressible FQH ground state in this limit.  A critical point separating the compressible and incompressible phases is  a possibility unrealized in the ideal limit of the FQHE. Although such a Mott transition in a Chern band could always be first-order, there is the additional intriguing possibility of  a {\it continuous} transition into FQH states. 

Several examples of such transitions at half-filling were studied by Barkeshli and McGreevy~\cite{PhysRevB.86.075136,Barkeshli12014393} and can be divided into three sets of phases which can have continuous transitions between each other:
\begin{enumerate}[(i)]
\item between superfluid, Mott insulating and $\nu=1/2$ Laughlin states of bosons;
\item between an ordinary Fermi liquid, a gapless Mott insulator, and  a composite fermion Fermi liquid  (i.e., the Chern band analog of the Halperin-Lee-Read state in the half-filled Landau level); and
\item between a $p+ip$ paired fermion superfluid, the B-phase of Kitaev's honeycomb model, and a lattice analog of the Moore Read Pfaffian phase.
\end{enumerate}
All these were studied within a parton mean-field theory which is expected to capture the gross features of the phases as well as the critical behavior between them. Their results may be summarized as follows. In essence, through the  parton decomposition, transitions between the different phases can be rewritten in terms of Chern number-changing transitions between insulators: $|\Delta C| = 1$ between successive members of each of the three sets above. A general critical theory for Chern-number changing transitions in two dimensions can be formulated in terms of massless Dirac fermions; a change in the sign of a Dirac mass changes the Chern number by $1$. In the cases studied by Barkeshli and McGreevy, the massless Dirac fermions are  also coupled to a Chern Simons gauge field, which emerges from the parton construction. For all three sets of phases, a {\it direct} transition between the first and the third phase requires $|\Delta C|=2$ and is generically multicritical, unless an additional symmetry, such as inversion, is present to force two Dirac fermions to undergo a simultaneous change in the sign of their masses.\footnote{An alternative possibility that gives $\Delta C=2$, a quadratic band-touching, is perturbatively unstable in $d=2$ and thus ruled out.} Furthermore, the parton construction describing (i) is the parent example; the remaining two  follow from a further parton decomposition in which an additional neutral fermionic degree of freedom is attached to each boson in (i) and allowed to form either a Fermi surface of gapless excitations (ii) or a paired state (iii). We direct the interested reader to Refs. \cite{Barkeshli12014393,PhysRevB.86.075136} for a detailed discussion of these transitions, along with a list of physical manifestations in crossover physics and scaling behavior.  

\section{Higher Chern Numbers and Nonabeliana}
Another possibility that can be realized in lattice models is a band with higher Chern number; the interest in these examples is primarily concerned with the possibility of realizing non-Abelian  statistics in a manner intrinsically tied to the lattice limit of the FQHE in a Chern band. In this context we note two intriguing directions we believe are particularly worthy of further attention. The first is a numerical study by Zhang and Vishwanath \cite{2012arXiv1209.2424Z}, who analyzed wave functions obtained by taking  $N$ copies of a  Slater determinant  for a fully filled Chern number $C$ band and Gutzwiller projecting to a Hilbert space that obeys a single-site-occupancy constraint.  A careful variational Monte Carlo study of the entanglement entropy of degenerate ground states on a torus for the case $N=C=2$ provides strong evidence that the resulting wavefunction has the same topological order as the Moore-Read Pfaffian state, described by an $SU(2)_2$ Chern-Simons theory. Combined with a previous analysis of  lattice analogs of bosonic Laughlin states along similar lines \cite{PhysRevB.84.075128}, this suggests that, in general, projecting the $N^{\text{th}}$ power of a Slater determinant of a filled Chern number $C=k$ band provides a lattice version of a  quantum Hall state whose  topological order is identical to that of the $SU(N)_k$ Chern-Simons theory coupled to fermions  (which also describes a system of $k$ `layers'  at total filling $\nu=k/N$.)

The interplay between the lattice-scale physics and topological order in a band with $C>1$ was studied in detail by Barkeshli and Qi \cite{PhysRevX.2.031013}. Their result can be readily understood within the Wannier description; recall that the center-of-mass position of a Wannier function shifts from $x \rightarrow x+C$ when $k_y$ evolves from $0$ to $2\pi$. If $C>1$, it is therefore possible to define $C$ distinct families of Wannier functions  that evolve independently under $k_y\rightarrow k_y+2\pi$. For  concreteness, consider the $C=2$ case. Here, the Wannier functions on even and odd sites form distinct families: in the notation introduced previously, we define 
\be
|{W^1_{K_y = k_y+2\pi n}}\rangle  &=&\ket{W(k_y, 2n-1)},\,\,\,\,\,\,
|{W^2_{K_y = k_y+2\pi n}}\rangle = \ket{W(k_y, 2n)}
\ee
On its own, the observation of two families of eigenfunctions is not particularly remarkable; a similar structure obtains, for instance, in multicomponent quantum Hall systems such as quantum Hall ferromagnets and bilayer systems. The key difference is the action of the lattice symmetry operations on the `internal' index. For instance, the translations  along $x$ interchange the internal index, which are invariant under translations along $y$:\footnote{Recent work \cite{2012arXiv1210.6356W} has shown that a slightly more involved definition of the Wannier orbitals and the associated action of the translation operator is required in order to impose boundary conditions that fully respect various lattice symmetries. Nevertheless, the interplay of lattice defects with topology and the field theory describing this structure remain, so we will ignore these subtleties.}
\be
\hat{T}_x |W^1_{K_y}\rangle =  |W^2_{K_y}\rangle,\,\,\,\,\,\, \hat{T}_x |W^2_{K_y}\rangle =  |W^1_{K_y}\rangle,\,\,\,\,\,\,
\hat{T}_y |W^a_{K_y}\rangle =  e^{i K_y}|W^a_{K_y}\rangle, a=1,2
\ee
As a consequence, lattice defects -- such as dislocations -- have a nontrivial impact on the underlying topological state. In a concrete example, Barkeshli and Qi \cite{PhysRevX.2.031013} considered Abelian `bilayer' FQHE states realized in a $C=2$ band, and showed that such defects that act nontrivially on the this internal `$\mathbb{Z}_2$' degree of freedom have a nontrivial topological degeneracy -- analogous to quasiparticle states in a {\it non}-Abelian state. In effect, by acting nontrivially on the layer index, these `twist' defects simulate the action of a change of genus of the underlying space -- and are hence dubbed `genons'. A subsequent detailed analysis  \cite{2012arXiv1208.4834B} of the projective non-trivial braiding statistics obtained in this fashion, while beyond the scope of the present article, suggests an interesting new line of enquiry with possible implications for topological quantum computation.

\section{Proposed Realizations}
Thus far our discussion has focused on lattice models and numerical results. As we emphasized earlier, an appealing feature of the lattice FQHE phases discussed above is that the characteristic energy scale at which new physics sets in is tied to the microscopic (lattice)  scale, which is quite large; if they can be experimentally realized, it is possible that Chern bands will exhibit fractionalization and associated phenomena at significantly higher temperature scales and be considerably more robust than in 2DEGs. We flag a few especially promising examples below.

In a solid-state context, Xiao and collaborators suggested that topological bands may be engineered in oxide interaces \cite{Xiao:2011fk}. From a combination of first-principles band structure calculations and tight-binding modeling, they predict that bilayer [111] heterostructures of transition-metal perovskite oxide host relatively flat bands with a nontrivial $\mathbb{Z}_2$ topological index. Apart from providing a  route to room-temperature quantum  spin Hall insulators, when the bands are split by a Zeeman term (this can be extrinsic, or in an interacting model emerge due to ferromagnetic ordering) they emerge with a non-zero Chern number. The flatness, which is due to the weak dispersion of the crystal-field split $d$-orbitals, suggests that fractional quantum Hall physics is a plausible scenario in these systems; see also \cite{PhysRevB.84.241103}.
  
Another direction is to directly engineer spin-orbit coupled models in optical lattices of ultracold atoms. Cooper \cite{PhysRevLett.106.175301} has shown that the strong-lattice limit can be achieved in two-species ultracold atom systems (where for instance the two species are the ground and long-lived excited states of an alkaline earth atom) effectively described by the single-particle Hamiltonian
\be
\hat{H}  = \left[\frac{\mathbf{p}^2}{2m} + V_0(\br)\right]\mathbf{1} +\mathbf{V}_s(\br) \cdot\boldsymbol{\sigma}
\ee
where the `spin' $\boldsymbol{\sigma}$ refers to the level index. Here, $V_0(\br)$ is a conventional scalar potential term, for instance from a conventional optical lattice, while $\mathbf{V}_s(\br)$ reflects a laser pattern that generates an `optical flux lattice', where the diagonal components reflect a species-dependent scalar potential, and the off-diagonal terms represent  a Raman coupling that tunes the rate of interspecies conversion as a function of position. A judicious choice of laser parameters $(V_0,\mathbf{V}_s)$ leads to a lowest band with a non-zero Chern number. It is also possible to adjust parameters such that the band is relatively flat \cite{2012arXiv1212.3552C}. This and similar proposals suggest that the high range of tunability afforded by optical lattice models will allow us to study a variety of features of Chern bands.

A different bosonic cold atom implementation is to consider the rotational degrees of freedom of ultracold molecules, whose actual spatial motion is pinned by a deep optical lattice. Specifically, Yao {\it et.~al.} \cite{2012arXiv1212.4839Y,PhysRevLett.109.266804} consider a system of such pinned, three-level dipoles in a two-dimensional lattices. A  fractional Chern insulator Hamiltonian can be shown to describe the spin flips of this problem which are effective hard core bosons. The time reversal breaking necessary for a nonzero Chern number is provided by elliptically polarized, spatially modulated electromagnetic radiation; the nontrivial spatial dependence  of the hopping relies on the the  long-range dipole-dipole interactions characteristic to polar molecules, which naturally introduces orientation-dependent hopping amplitudes. Appropriately tuning these also allows control of the hopping interference and permits bands  with $f\gtrsim 10$ \cite{PhysRevLett.109.266804}. Numerical studies of these models show a variety of phases at $\nu=1/2$, including fractional Chern insulators for a range of parameters that are not too far from  current experimental capabilities  \cite{2012arXiv1212.4839Y}.

\section{Concluding Remarks}
We hope that the reader who has persevered to this point has acquired an impressionistic perspective of several quite different aspects of this rather active area of research. We have taken a more or less historical approach, first explaining how to design topological flat band models and providing an overview of the numerical results, before drawing analytical parallels between fractional Chern insulators and FQH physics in the lowest Landau level. We then briefly touched upon a host of interesting topics:  constructing  and classifying fractional Chern insulator wave functions using parton mean-field theories, the theory of bandwidth-tuned transitions into FQH phases, and various aspects of lattice physics in bands of higher Chern number. Each of these has spawned its own numerical and analytical offshoots, and we have attempted to provide pointers to the relevant  literature wherever possible. We also gave an overview of recent progress in developing experimentally realistic scenarios where these correlated phases emerge.

In closing this review, we take a somewhat longer view, and outline several directions which seem (to us, at least) worthy of further study, although we hope that the preceding discussion has already prompted our readers to form similar questions independent of these. Once again, we  sound a note of caution: given the rapid progress in the field, it is quite likely that this list of future directions might quite rapidly become dated as new results emerge on various fronts.

{\it Disorder.---} The first of these is the important and  (at this point) relatively little-studied problem of disorder in the Chern band. Here it is worth noting that in addition to the square root of the mean Berry curvature (which as we have seen plays a role analogous to the magnetic length in the LLL), we have a second (inverse) length scale: the characteristic momentum-space scale $k_\sigma$ for variations of the Berry curvature. Na\"ively we might expect that the role of disorder depends on how its strength -- as quantified by the mean free path, $\l_\text{MF}$ -- compares to one or both of these scales.  For instance, in a problem with sufficient disorder, most notably the problem of localization within the CB, we expect that impurity scattering will effectively lead to an averaging of the curvature over the band and give results similar to those in the LLL. This is consistent with what is known about the (integer) QH transition in Chern insulators \cite{Onoda:2003p1} although a direct test in the projected CB would be desirable. A simple estimate for the requisite strength of disorder can be made by comparing $l_\text{MF}$ to $k_\sigma^{-1}$;  for disorder sufficiently strong that $l_\text{MF}\sim k_\sigma^{-1}$, the random potential will scatter between points in the BZ that are $k_\sigma$ apart, and thereby average their Berry curvature. More ambitiously, it would be quite instructive to construct a global phase diagram for fractional Chern insulators, akin to that proposed by Kivelson, Lee and Zhang \cite{Kivelson:1992p1550} for the FQHE in the LLL. Here specifically it is worth recalling that  the full global  phase diagram for the continuum Landau problem emerges only when mixing between different LLs is taken into account; for fractional Chern bands it is possible that the mixing between Bloch bands leads to qualitatively different physics.

{\it Long-Range Interactions and Competing Phases.---} Turning for a moment to the continuum problem, we observe that there fractionalized phases emerge in a comparatively narrow window of parameter space: they compete with integer QH liquids as well as  electronic crystalline phases, the latter in particular drawing enhanced stability from long-range Coulomb physics.  We note that so far numerical studies in the Chern band have been largely confined to Hamiltonians with relatively short-ranged interactions. For the FQHE this is typically justified by invoking screening to argue that the Coulomb repulsion is operational over no more than a few magnetic lengths -- although we note that even in the LLL there are well-known examples where the long-range tail of the Coulomb interactions carries a sting \cite{Sondhi:1992p1220}. In a CB in contrast, the only length scale is the inter-atomic spacing, and even a screened interaction can be quite long-ranged on the lattice scale. The role of such longer-ranged interactions in stabilizing or inhibiting the formation of FQH phases, and in particular in supporting competing crystalline phases, may be quite relevant to experimental systems. We note that some numerical work has begun to explore competition between phases in the Chern band \cite{2012arXiv1207.6094L}.

{\it Edge States.---} An integral aspect of the theory of the FQHE is the Luttinger liquid description of the gapless edge modes \cite{Wen:1990p3}. The structure of the edge of a fractional Chern insulator has received comparably little attention. While presumably the universal edge properties are captured by the standard chiral Luttinger liquid theory, it is possible that the ability to produce a lattice-scale edge and the interplay of lattice symmetries with the edge may provide a source of additional complexity, worth exploring.

{\it Experimental Detection.---} As we have described already, experiments on fractional Chern insulators remain in their infancy, although several proposals seem quite compelling and suggest that a realization even with existing technology is not too far off. In this context it is important to develop techniques to characterize the microscopic properties of topological band structure, which as we have seen play an important role in stabilizing correlated phases. In light of this, a recent proposal \cite{2012arXiv1212.0562A} to measure the Zak phase in Bloch bands using a combination of Ramsey intereference and Bloch oscillations are promising, and have been demonstrated in experiments on 1D optical lattices \cite{2012arXiv1212.0572A}. Equally important is the development of methods to {\it identify} correlated topological phases, particularly since the presence of the lattice leads to ambiguities in interpreting standard measures such as the Hall conductance.

{\it Matrix Product and Tensor Network Representations.---} The study of gapped phases in one dimension has benefited greatly from the observation that they have ground states that can be efficiently represented as `matrix product states' \cite{1992CMaPh.144..443F}.  
The power of the matrix product form is twofold: first, the density-matrix renormalization group (DMRG) algorithm \cite{PhysRevLett.69.2863,PhysRevLett.75.3537} efficiently obtains the best matrix product representation of a ground state of a given local Hamiltonian; and second, the structure of the MPS allows efficient analysis of its behavior under symmetry operations. Similar tensor network constructions exist in higher dimensions \cite{2004cond.mat..7066V} but a full understanding of these is still being built.  The FQH phases are gapped correlated phases, and recently several groups have reformulated the model wavefunctions as matrix product states \cite{PhysRevB.86.245305,2012arXiv1211.3353E}. It would be quite instructive to develop a similar understanding for FQH phases in Chern bands, as well as to develop efficient algorithms to simulate them.
 
{\it Extensions.---} We turn finally to the important and interesting question of what other correlated topological phases might emerge in lattice models. One obvious extension that has already received quite some numerical and analytical attention is the problem of constructing interacting analogs of time-reversal invariant topological insulators in two and three dimensions. Unlike the FQHE, this is a problem with little or no prior history, so there are many open questions; some work along these directions already exists \cite{PhysRevLett.103.196803,PhysRevLett.105.246809,PhysRevB.83.195139,PhysRevB.84.165107,PhysRevB.84.165138,PhysRevB.84.235145}.
A more recent development is the classification of `symmetry-protected topological phases' (SPTs) \cite{2013arXiv1301.0861C}. These are gapped phases which have no topological order (they lack ground-state degeneracy and fractionalized excitations) but still host gapless edge states, and cannot be adiabatically continued to a trivial gapped phase as long as a certain symmetry is left unbroken -- for instance, topological band insulators of non-interacting fermions may be thought of as SPTs protected by time-reversal symmetry, while the Haldane phase of a spin-$1$ chain is protected by $\pi$-spin rotation, time reversal, or inversion symmetry\cite{PhysRevB.81.064439,PhysRevB.85.075125}
.\footnote{Note that while SPTs are not topologically ordered, they may still be correlated, as exemplified by bosonic SPT insulators.} Since the understanding of SPTs grew out of very similar considerations as the integer and fractional QHE, it is quite natural to consider Chern bands as a natural venue where they might be realized. A final note of complexity emerges when one considers the interplay of lattice symmetries and topological order, as in a few examples above; work in this direction has potential to reveal signatures of hard-to-detect topological phenomena in more conventional probes of the solid state. 

\section{Acknowledgements}
Our understanding of various topics discussed here has benefited greatly from discussions with several colleagues, and it would be impossible to thank them each individually. However, we would be remiss if we did not acknowledge B. A. Bernevig, N. Regnault, S. H. Simon, X.-L. Qi, M. Barkeshli, R. Thomale, R. Shankar, G. Murthy, G. M\"oller, M. Zaletel, and F. D. M. Haldane for several conversations and patient explanations of their work. In addition, SLS thanks B. A. Bernevig for numerous insightful discussions on the feasibility of studying correlated phases emerging from Hamiltonians projected into topological bands some time prior to the recent explosion of work, which inspired our own work \cite{parameswaran2012fractional}. We also thank the the authors of Refs. \cite{Neupert:2011p1,sheng2011fractional,regnault2011fractional} for kindly providing us with figures of their numerical results. We  acknowledge support from the Simons Foundation (SAP) and the  NSF  through grants DMR-1006608 and PHY-1005429 (SAP, SLS), the  EPSRC through grant  EP/D050952/1 and UCLA Startup Funds (RR).
\bibliographystyle{elsarticle-num}
\bibliography{FractionalCI_bib}
\end{document}